\documentclass[twocolumn,times,tighten,twocolappendix]{aastex63}

\newcommand\aastex{AAS\TeX}

\newcommand{\msun}{\mathrm{M}_{\sun}}

\newcommand{\ha}{H$\alpha$}
\newcommand{\hb}{H$\beta$}
\newcommand{\hg}{H$\gamma$}
\newcommand{\hd}{H$\delta$}

\newcommand{\oiii}{[O{\sc III}]5007}
\newcommand{\oii}{[O{\sc II}]3726,3729}

\newcommand{\vrot}{$v_{rot,R_e}$}
\newcommand{\sig}{$\sigma_0$}
\newcommand{\vc}{$v_{c,R_e}$}

\newcommand{\mgas}{$\mathrm{log \left[M_{dyn,eml}/\msun\right]}$}
\newcommand{\mjam}{$\mathrm{log \left[M_{dyn,*}/\msun\right]}$}

\newcommand{\dm}{$\mathrm{log \left[M_{dyn,eml}/M_{dyn,*}\right]}$}

\newcommand{\ns}{$n_{S\acute{e}rsic}$}

\newcommand{\finalsample}{157}
\newcommand{\nfinalsample}{$\mathrm{N}=$\finalsample}

\newcommand{\mainoffset}{$-0.15\pm0.016$}
\newcommand{\mainabsoffset}{$0.15\pm0.016$}
\newcommand{\mainscatter}{$0.19$}

\usepackage{ulem,amsmath,xcolor,soul}

\shorttitle{\aastex\ Dynamical masses at $z\leq1$}
\shortauthors{Straatman et al.}

\begin{document}

\title{LEGA-C: analysis of dynamical masses from ionized gas and stellar kinematics at $z\sim0.8$}

\correspondingauthor{Caroline M. S. Straatman}
\email{Caroline.Straatman@Ugent.be}

\author[0000-0001-5937-4590]{Caroline M. S. Straatman}
\affil{Department of Physics and Astronomy, Ghent University, Krijgslaan 281 S9, B-9000 Gent, Belgium}

\author[0000-0002-5027-0135]{Arjen van der Wel}
\affil{Department of Physics and Astronomy, Ghent University, Krijgslaan 281 S9, B-9000 Gent, Belgium}
\affil{Max-Planck Institut f\"ur Astronomie, K\"onigstuhl 17, D-69117, Heidelberg, Germany}

\author{Josha van Houdt}
\affil{Max-Planck Institut f\"ur Astronomie, K\"onigstuhl 17, D-69117, Heidelberg, Germany}

\author[0000-0001-5063-8254]{Rachel Bezanson}
\affil{University of Pittsburgh, Department of Physics and Astronomy, 100 Allen Hall, 3941 O’Hara St, Pittsburgh PA 15260, USA}

\author[0000-0002-5564-9873]{Eric F. Bell}
\affil{Department of Astronomy, University of Michigan, 1085 S. University Ave., Ann Arbor, MI, 48109, USA}

\author[0000-0002-8282-9888]{Pieter van Dokkum}
\affil{Astronomy Department, Yale University, New Haven, CT 06511, USA}

\author[0000-0003-2388-8172]{Francesco D'Eugenio}
\affil{Department of Physics and Astronomy, Ghent University, Krijgslaan 281 S9, B-9000 Gent, Belgium}
\affil{Cavendish Laboratory and Kavli Institute for Cosmology, University of Cambridge, Madingley Rise, Cambridge, CB3 0HA, United Kingdom}

\author[0000-0002-8871-3026]{Marijn Franx}
\affil{Leiden Observatory, Leiden University, P.O. Box 9513, 2300 RA, Leiden, The Netherlands}

\author[0000-0002-9656-1800]{Anna Gallazzi}
\affil{INAF-Osservatorio Astrofisico di Arcetri, Largo Enrico Fermi 5, I-50125 Firenze, Italy}

\author[0000-0002-2380-9801]{Anna de Graaff}
\affil{Leiden Observatory, Leiden University, P.O. Box 9513, 2300 RA, Leiden, The Netherlands}

\author[0000-0003-0695-4414]{Michael Maseda}
\affil{Leiden Observatory, Leiden University, P.O. Box 9513, 2300 RA, Leiden, The Netherlands}

\author[0000-0002-6118-4048]{Sharon E. Meidt}
\affil{Department of Physics and Astronomy, Ghent University, Krijgslaan 281 S9, B-9000 Gent, Belgium}

\author[0000-0002-9330-9108]{Adam Muzzin}
\affil{Department of Physics and Astronomy, York University, 4700 Keele St., Toronto, Ontario, M3J 1P3, Canada}

\author[0000-0001-8823-4845]{David Sobral}
\affil{Department of Physics, Lancaster University, Lancaster LA1 4YB, UK}

\author[0000-0002-9665-0440]{Po-Feng Wu}
\affil{Institute of Astronomy and Astrophysics, Academia Sinica, No.1, Sec. 4, Roosevelt Road, Taipei 10617, Taiwan, R.O.C.}

\begin{abstract}

We compare dynamical mass estimates based on spatially extended stellar and ionized gas kinematics ($\mathrm{M_{dyn,*}}$ and $\mathrm{M_{dyn,eml}}$, respectively) of \finalsample\ star forming galaxies at $0.6\leq z<1$. Compared to $z\sim0$, these galaxies have enhanced star formation rates, with stellar feedback likely affecting the dynamics of the gas. We use LEGA-C DR3, the highest redshift dataset providing sufficiently deep measurements of a $K_s-$band limited sample. For $\mathrm{M_{dyn,*}}$ we use Jeans Anisotropic Multi-Gaussian Expansion models. For $\mathrm{M_{dyn,eml}}$ we first fit a custom model of a rotating exponential disk with uniform dispersion, whose light is projected through a slit and corrected for beam smearing. {We then apply an asymmetric drift correction {based on assumptions common in the literature }to the fitted kinematic components to obtain the circular velocity, assuming hydrostatic equilibrium. }Within the half-light radius, $\mathrm{M_{dyn,eml}}$ is on average lower than $\mathrm{M_{dyn,*}}$, with a mean offset of \mainoffset\ dex and galaxy-to-galaxy scatter of \mainscatter\ dex, reflecting the combined random uncertainty. {While data of higher spatial resolution are needed to understand this small offset, it }supports the assumption that the {galaxy-wide} ionized gas kinematics do not predominantly originate from disruptive events such as {star formation driven} outflows. 
{However, }a similar {agreement} can be obtained {without modeling }from the integrated emission line dispersions for axis ratios $q<0.8$. {This suggests that our current understanding of gas kinematics is not sufficient to efficiently apply asymmetric drift corrections to improve dynamical mass estimates compared to }observations lacking the $S/N$ required for spatially {extended} dynamics.

\end{abstract}

\keywords{galaxies: evolution --- galaxies: high-redshift --- galaxies: kinematics and dynamics}

\section{Introduction} \label{sec:intro}

The dynamical mass of a galaxy is the total mass enclosed within a specific radius as implied by {orbital motions of an observable tracer, such as stars}{, and as such one of its most fundamental properties. As new generations of telescopes open new windows onto the universe, probing higher redshifts or resulting in more detailed measurements, it is important to understand how different tracers of dynamical mass relate to each other, before drawing conclusions on the mass properties of galaxies themselves. The aim of this study is to compare dynamical mass estimates based on the motions of stars with those based on ionised gas motions for galaxies at $0.6<z<1$.}

Dynamical mass can be measured with a variety of tracers. For the Milky Way the {orbital motions} can be measured out to tens of kpc from the motions of individual stars \citep[e.g.,][]{Eilers19}. Beyond the Milky Way, arguably the most powerful tool to probe the dynamical mass profile as well as the dark matter fraction {of galaxies} is the 21 cm line of {neutral }Hydrogen. {This type of measurement is only available for studies of the nearby universe.} In studies of high redshift galaxies, dynamics are almost exclusively measured from {strong} emission lines, most notably from \ha$+$[N{\sc II}], [O{\sc II}] or [O{\sc III}], probing the region inside $1-2$ half-light radii. These lines {are due to warm ionized gas and }can be measured even when the stellar continuum is too faint to observe. However, the more reliable tracers of dynamical mass are the stellar atomic lines, because, albeit intrinsically complex, unlike gas particles, the motions of stars are not affected by local disruptive events, such as stellar feedback. Additionally, the ionized gas accounts for only a small fraction of the total amount of gas inside galaxies and its distribution is not necessarily representative of the total mass distribution{, or kinematically coupled to the other gas components}.

The intricacies associated with star formation in galaxies make the interpretation of {ionized gas} as a tracer of potential well complicated{, especially for galaxies with large observed velocity dispersions}. {In recent years, large spectroscopic high-redshift surveys, e.g., DEEP2 \citep{Newman13}, VUDS \citep{LeFevre15}, MOSDEF \citep{Kriek15}, KMOS$\mathrm{^{3D}}$ \citep{Wisnioski19}, and KROSS \citep{Stott16}, have capitalized on emission lines originating from ionized gas to better understand the dynamics of thousands of galaxies beyond $z>0.6$. }One notable result of these surveys is that, at any fixed stellar mass, the galaxy-wide velocity dispersion of the ionized gas {in star forming }galaxies has progressively decreased over time since at least $z\sim3$, whereas the rotational velocities have increased \citep{Epinat10,Gnerucci11,Epinat12,Kassin12,Vergani12,Law09,Green14,NFS14,Wisnioski15,Turner17,Johnson18,Ubler19}. This phenomenon includes a downsizing effect \citep{Kassin12}{, i.e., {gas in }more massive galaxies tends to become kinematically dominated by rotation at earlier times {than in lower mass galaxies}, and on average has a higher ratio between rotation and dispersion than {gas in} lower mass galaxies at {the same epoch}}. Because of this downsizing effect, the fraction of rotation dominated, or dynamically ``settled'' disks \citep{Kassin12} was smaller at high redshift. Another consequence is that,{ as the average dispersion used to be much higher, dynamical mass estimates require an asymmetric drift correction to take into account the non-negligible dynamical support from velocity dispersion.} 

The velocity dispersion may reflect motions that are due to turbulence generated by SNe feedback \citep{Dib06}, stellar winds and/or radiation pressure from OB stars \citep[][and references therein]{MacLow04}. {In the scenario of self-regulated star formation \citep[e.g.,][]{Ostriker10,Santini17}, these feedback-driven motions maintain the disk in hydrostatic equilibrium. Galaxy-wide gravitational processes,} such as clump-clump interactions \citep{Dekel09,Ceverino10}, cold gas accretion \citep{Aumer10,Elmegreen10}, disk instabilities \citep{Immeli04,Bournaud10} and minor mergers \citep{Bournaud09}, {will also contribute to the observed velocity dispersion in multi-phase gas.  In equilibrium, turbulent pressure balances the weight of the rotating gas disk in the background gravitational potential regardless of whether this is a contribution from stellar feedback \citep{Meidt18,Meidt20}. The velocity dispersion in the ionized gas in equilibrium would then need to be accounted for with an asymmetric drift correction to the rotational velocity when attempting to trace the potential. {In this paper, we refer to all such motions as ``gravitational''.} But when the non-rotational motions are in the form of, e.g., outflows, their contribution to the velocity dispersion should not necessarily be expected to obey equilibrium and }should not be included in a dynamical mass estimate, unless {the total} can be decomposed into various equilibrium and non-equilibrium dynamical components. {This is nearly impossible at high redshift on sub-kpc scales, however for a first attempt using the galaxy wide velocity dispersion see, e.g., \citet{Wisnioski18}}. In this paper, we {refer to such non-equilibrium motions as ``non-gravitational''.}

{At low redshift, the kinematic properties of ionized gas disks, i.e., their rotational velocities and velocity dispersions are consistent with an equilibrium scenario \citep{Levy18}.  At high redshift, the enhanced gas fractions and star formation rates have been associated with observations of strong outflow signatures in the ionized gas \citep[e.g.,][]{Rubin10,NFS19}.  Thus we might expect gas velocity dispersions at earlier epochs to reflect an increased non-gravitational component, although it remains a question to what extent this is detectable in galaxy-wide measures of the velocity dispersion typically observed for high-redshift surveys.}

{Comparison to the dynamics of the underlying (collisionless) stellar disk is a powerful way to recognize whether the ionized gas motions are gravitational or non-gravitational in nature.} \citet{CrespoGomez21} {investigated the dynamics of local luminous infrared galaxies (LIRGs), which can be considered local counterparts of the star forming galaxy population at $z > 1$. This study found good correspondence between dynamical masses derived from ionized gas, molecular gas, and stars,} 
suggesting that the ionized gas kinematics is also an accurate tracer of dynamical mass at high redshift.

In this paper we compare dynamical mass estimates at $0.6<z<1$, based on the motions of ionized gas and those measured from the spatially {extended} dynamical features within the stellar continuum and absorption lines. We use the Large Early Galaxy Astrophysics Census \citep[LEGA-C;][]{vdWel16}, which is the first survey combining sufficient depth and sample size to allow measurements of stellar dynamics of hundreds of galaxies at $0.6<z<1$. The survey also comprises high $S/N$ measurements of \hb, \hg, \hd, \oiii, and \oii. The goals of this paper are 1) to cross-calibrate the two widely used tracers, ionized gas and stellar continuum/absorption, for the first time for galaxies at large lookback time, and 2) to test the assumption inherent to applying asymmetric drift corrections that kinematics seen in ionized gas {reflect some form of equilibrium}. {We note that the limited spatial resolution of high redshift studies makes it difficult to draw more direct conclusions about the origins of gas turbulence.}

In Section \ref{sec:data} we describe the survey and in Section \ref{sec:sample} the sample selection. In Section \ref{sec:method} {we describe the methodology used to extract dynamical masses from the motions of stars and ionized gas. In Section \ref{sec:results} we compare these different dynamical mass estimates. In Section \ref{sec:discussion} we show residual trends with other galaxy properties and we discuss our results in the context of a gravitational versus non-gravitational origin of the ionized gas motions, as well as possible caveats. }
In Section \ref{sec:summary} we provide a summary of this work.

Throughout, we assume a standard $\Lambda$CDM cosmology with $\Omega_{\mathrm{M}}=0.3$, $\Omega_{\Lambda}=0.7$, and $H_0=70\ \mathrm{km\ s^{-1}\ Mpc^{-1}}$, unless otherwise mentioned.

\section{Data}\label{sec:data}

We used data from LEGA-C Data Release 3 \citep[DR3;][]{vdWel16,vdWel21}. LEGA-C is an ESO 130-night public spectroscopic survey of COSMOS conducted with VIMOS \citep{LeFevre03} on the Very Large Telescope. 4209 exceptionally deep, spatially {extended} {slit }spectra were taken of 3855 unique sources in the UltraVISTA photometric catalog \citep{Muzzin13}, primarly chosen from the redshift range $0.6\lesssim z \lesssim 1$. Typical integration times were 20 hrs. {The resolution properties of the LEGA-C spectra are $R\sim3500$ and FWHM$=86\mathrm{km\ s^{-1}}$ ($\sigma=36\mathrm{km\ s^{-1}}$).} The typical {integrated }$S/N$ of {the continuum of }a spectrum is $20\mathrm{\AA}^{-1}$. 

{Primary targets for LEGA-C are $K_s-$band selected, with limits ranging from $K_s<21.08$ at $z=0.6$ to $K_s<20.36$ at $z=1.0$. These limits are equivalent to a stellar mass limit of $\sim10^{10}\msun$. We applied a two-step data reduction, using first the ESO and then a custom built LEGA-C pipeline. The custom pipeline was needed {to improve} sky-subtraction and optimal $S/N-$weighted extraction of spectra. The reduction process was described in detail in \citet{Straatman18} and was updated for DR3 in \citet{vdWel21}. In total the survey comprises {2942} primary targets with successful redshift measurements.}

For each individual spectrum, each row with median $S/N>2$ per pixel was fit with Penalized Pixel-Fitting (pPXF) code \citep{Cappellari04,Cappellari17}, using two template sets. These consisted 1) of a collection of high resolution ($\mathrm{R}=10\ 000$) single stellar population templates, downgraded to match the resolution of the LEGA-C spectra, {with solar metallicity, a Salpeter IMF and ages in the range $10^{-4} - 10^{1.3}$ Gyr, }to fit the continuum (Conroy et al., in prep.), as well as  2) a collection of possible emission lines ([Ne{\sc V}], [Ne{\sc VI}], H10, H9, H8, \hd, \hg, \hb, [O{\sc II}]3726,3729, [Ne{\sc III}], [S{\sc II}]6717,6731, [O{\sc III}]5007, [O{\sc III}]4959, [O{\sc III}]4363, [N{\sc I}]) to fit the ionized gas emission. The emission line fluxes were allowed to vary independently from each other. The linewidths were allowed to vary between rows in the spectra, but were constrained to be the same for all lines at different wavelengths in a single row \citep[see also][]{vdWel16,Bezanson18a,Bezanson18b}.  In this way we obtained one dynamical measurement per row. We derived raw rotational velocities ($v_{rot,pPXF}$) and dispersions $\sigma_{pPXF}$ from the {line centroids} and {gaussian }line widths of the best fits, with the Conroy et al. (in prep) templates resulting in constraints on the stellar dynamics and the emission line collection in constraints on the ionized gas dynamics. 

Stellar masses and (specific) star formation rates ((s)SFRs) were obtained from the UltraVISTA photometry by running the Bayesian inference of star formation histories code Prospector \citep{Leja17,Johnson19}, which includes the Affine Invariant Markov chain Monte Carlo (MCMC) Ensemble sampler of \citet{ForemanMackey13}{, using the spectroscopic redshifts from LEGA-C}. {The photometric bands used were the Subaru/SuprimeCam $r^+,i^+,z^+,B_j-$ and $V_j-$ bands \citep{Capak07} and the VISTA $Y-$ and $J-$bands \citep{McCracken12}, complemented with $24\mu m$ data from Spitzer/MIPS (PI: Scoville), with adjusted zeropoints \citep[see][for details]{vdWel21}.} The Prospector setup was identical to the one used by \citet{Leja19}, including a \citet{Chabrier03} IMF and a standard cosmology ($\Omega_{\Lambda}=0.7$, $\Omega_{m}=0.3$, $\mathrm{H_0=70\ km\ s^{-1}\ Mpc^{-1}}$).

{A total of }$93\%$ of LEGA-C targets are covered by the COSMOS HST/ACS/F814W mosaic \citep{Scoville07}. We used GALFIT \citep{Peng10} to extract S\'ersic profiles of each source. The resulting parameters: half-light radius along the major axis ($R_{e}$), S\'ersic index (\ns), magnitude, position angle (PA) and axis ratio ($q=b/a$), are included in the DR3 catalog.

\section{Sample selection}\label{sec:sample}

\begin{figure*}
\centering
\includegraphics[width=0.8\textwidth]{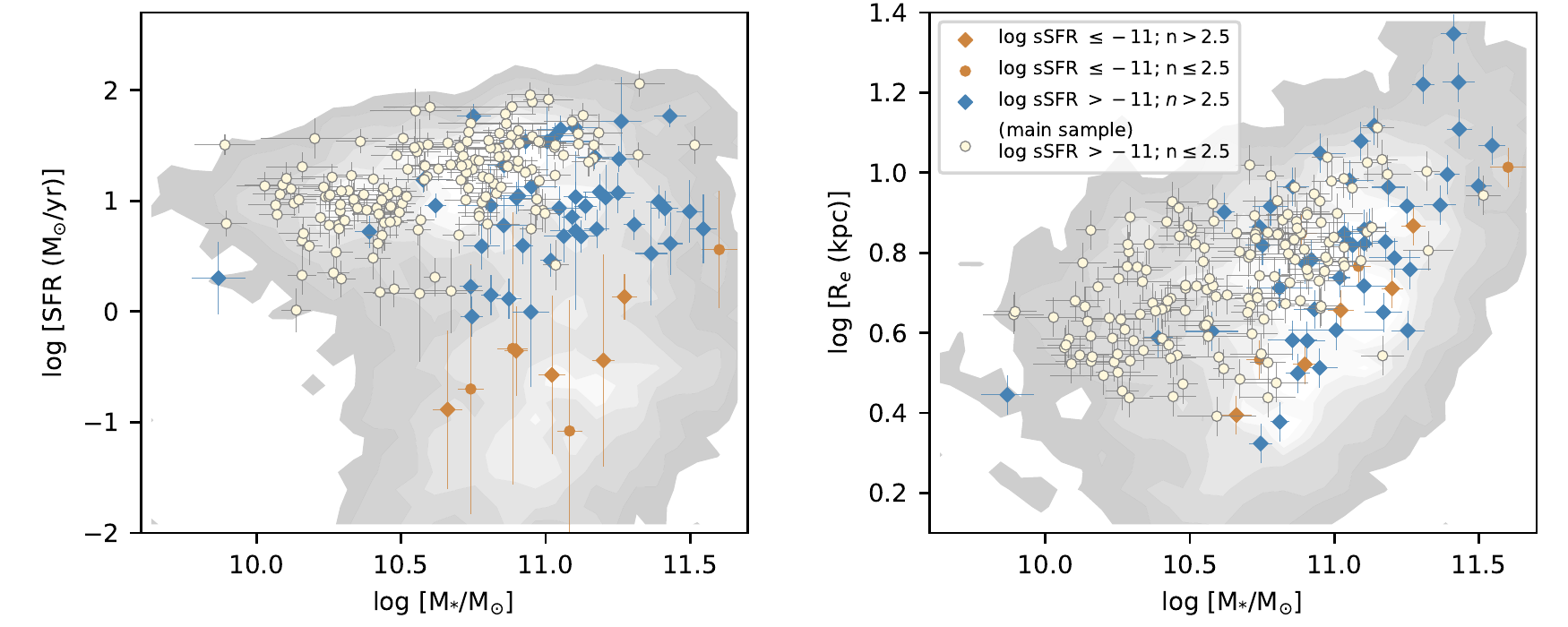}
\caption{\label{fig:fig1} {Left: SFR versus stellar mass distribution of the LEGA-C parent sample (gray contours) and main sample used in this study (yellow dots). Right: half-light radius versus stellar mass with symbols as in the left panel. In addition to the main sample (N$=$\finalsample; $\mathrm{log\ sSFR (yr^{-1})}>-11$; \ns$\leq2.5$) we show high S\'ersic index star forming galaxies (blue diamonds), as well as quiescent galaxies with \ns$>2.5$ (orange diamonds) and \ns$\leq2.5$ (orange dots). Due to a lack of detected Balmer emission lines, the quiescent subsample is highly incomplete. The higher S\'ersic index star forming galaxies tend to have lower SFRs than the main sample and include the most massive and largest galaxies of all galaxies that pass the combined criteria. Whereas the LEGA-C survey is close to complete above $>10^{10}\msun$, due to the requirement of ``resolved'' rotation curves, 
the main sample here could be incomplete near that limit.}}
\end{figure*}

{Our parent sample consists of all {2608\footnote{There are less primary targets with spectroscopic redshifts (2608) than with photometric redshifts (2942) at $0.6\leq z<1$.}} primary LEGA-C targets with successful spectroscopic redshifts (\texttt{Z\_SPEC}$>0$) at $0.6\leq z<1$.  The stellar mass versus star formation rate and effective radius distributions of the parent sample are shown with contours in Figure \ref{fig:fig1}. To this sample we applied a number of cuts, mostly driven by the N$-$S orientation of the slits, to obtain those sources for which we can reliably derive dynamical masses. We note that we will only use a fraction of the whole survey, as the survey strategy was not optimized to study spatially extended dynamics in star forming galaxies. }

The first cuts were based on dynamical arguments. We cannot detect rotation in face-on galaxies, so we employed a $q=0.9$ cut-off. We furthermore selected only those galaxies whose major axes are aligned with the N$-$S slits to within $\Delta\mathrm{PA}<|45^{\circ}|$. {This cut-off is similar to other works in literature \citep[e.g.,][]{Weiner06}. We verified that the formal uncertainties on our results are independent of $|\Delta\mathrm{PA}|$. We also verified that the ratio between dynamical mass (Section \ref{sec:method}) and stellar mass, with the latter determined from photometry and thus unrelated to slit angle, has no trend with $|\Delta\mathrm{PA}|$. }We note that the alignment between the gaseous and stellar components of galaxies can be different: \citet{Wisnioski15} {found 40\% and 20\% of KMOS$\mathrm{^{3D}}$ galaxies at $0.7<z<2.7$ have such different alignments by $>15^{\circ}$ and $>30^{\circ}$ respectively, with the chance of misalignment highest for more face-on orientations, and \citet{Harrison17} found an average misalignment of $13^{\circ}$ for rotationally dominated galaxies at $0.6<z<1$}, but this can only be determined with IFU data. Given the limitations of single-slit spectroscopy, we assume the stellar PA is the same as that of the gas and accept that potential misalignments could {potentially contribute to the scatter in} our results. {The limitations of our data mean that this contribution is otherwise unquantifiable within the scope of this work.} Finally, we required that the dynamical data are ``resolved'', i.e., that the stellar $R_e$ as measured by GALFIT, as well as the extent of the rotation curves exceeds $0.4\times$ the PSF FWHM. These steps reduced the sample to {1011} sources {and automatically excluded the 7\% of LEGA-C targets outside of the COSMOS HST/ACS/F814W footprint and without data on $q$, PA, and $R_e$.}

{The sample was further restricted based on morphology. We excluded galaxies with \texttt{FLAG\_MORPH}$>0$, which are galaxies with irregular morphologies such as merger remnants, multiple galaxies not separated in the spectrum, or galaxies that suffer contamination from a neighbour at a different redshift. We also excluded galaxies with \texttt{SERSIC\_LIMIT}$>0$, i.e., those for which the GALFIT parameters reached their limiting values, e.g., $n_{S\acute{e}rsic}=0.2$ or $n_{S\acute{e}rsic}=6$.} %
{With these steps we removed a total of {194} additional galaxies from the sample.}%

The final cuts were based on spectral quality: we selected sources with a continuum $S/N>10\mathrm{\AA}^{-1}$, and with at least $3\sigma$ \hb\ or \hg\ detections. The \hb, \hg, \oiii, and \oii\ lines were visually inspected and sources were removed that do not have enough coverage between these four lines to be free of significant contamination by skylines. We also removed sources for which the ionized gas emission is clearly dominated by \oii, but where the doublet line ratio is not well determined by pPXF and (part of) the rotation curve is offset to one of the two lines of the doublet. {We did not remove sources that appear to have an irregular or clumpy light distribution. Finally, we removed AGN from the sample (sources with \texttt{FLAG\_SPEC}$>0$), as in some cases AGN affected the continuum shape and/or the flux calibration of the LEGA-C spectra \citep[for further details see][]{vdWel21}. This affects but a small number: there are only {85} ($3\%$) AGN in the parent sample.}

{The combined cuts resulted in a sample of \textbf{221} unique sources, which can be further split into star forming ($\mathrm{log\ sSFR (yr^{-1})}>-11$) or quiescent ($\mathrm{log\ sSFR (yr^{-1})}\leq11$) and high (\ns$>2.5$) or low (\ns$\leq2.5$) S\'ersic index galaxies; Figure \ref{fig:fig1}). Our main sample consists only of star forming, low S\'ersic index galaxies ($\mathrm{log\ sSFR (yr^{-1})}>-11$; \ns$\leq2.5$), comprising \nfinalsample\ unique galaxies. {The S\'ersic index cut-off was applied for consistency with our emission line dynamical mass model (Section \ref{sec:method_gas}), for which we assume exponential light profiles. The main sample is furthermore the most representative of the star formation sequence, as can be seen in Figure \ref{fig:fig1}, and less prone to suffer from potential quenching effects.} 7 galaxies are duplicates, i.e., they are included in more than one mask. For duplicates we always take a weighted mean of the calculated dynamical quantities.}

For both the stars and the ionized gas{ in the main sample}, the typical extent of the rotation curves is $1.5R_e$. For $5\%$ of the sources the gas dynamical masses within $R_e$ are {based on }an extrapolation of $>20\%$ of {best-fit models (see next Section) to }the data. This increases to $75\%$ for $2R_e$. For the stellar dynamical masses this is $4\%$ and $68\%$. The statistics mentioned in the remainder of this paper are for 149 ($95\%$) sources {with extended rotation curves ($<20\%$ extrapolation)} at $R_e$ and 40 ($25\%$) sources {with extended rotation curves} at $2R_e$ {for both the gas and the stars.}

\section{Dynamical modeling techniques}\label{sec:method}

We employ different methods to model the stellar and gas kinematics. For the stellar light we use the Jeans Anisotropic MGE models \citep[JAM][]{Cappellari02,Cappellari08}. For full details of the JAM analysis, we refer to the work of \citet{Cappellari02,Cappellari08} and for the LEGA-C specific implementation to \citet{vHoudt21}, but we give a summary in Section \ref{sec:jam}. For the emission lines we developed a custom model (Section \ref{sec:HELA}).%

The main focus of this paper is the analysis of the dynamics of ionized gas. We discuss methodology in Section \ref{sec:HELA}, results in Section \ref{sec:hela_results}, and  uncertainties in Section \ref{sec:eml_uncertainties}.

\subsection{Stellar dynamical masses}\label{sec:jam}

The ``stellar dynamical masses'' are the dynamical mass estimates based on the symmetrized absorption line based dynamics of the LEGA-C spectra \citep[for details, see][]{vHoudt21}. They were obtained by fitting Jeans Anisotropic Multi Gaussian Expansion models to the stellar rms velocity, based on the stellar line-of-sight dispersions and rotational velocities as measured with pPXF: $v_{rms,*}=\sqrt{v_{rot,*,pPXF}^2+\sigma_{*,pPXF}^2}$. The JAM models {were} based on the Jeans equations, which describe the motions of stars in a gravitational field and take the Collisionless Boltzmann Equation as a starting point. The Jeans equations {were} simplified assuming an axisymmetric shape, {a Gaussian velocity distribution, }alignment of the velocity ellipsoid with the main cylindrical coordinate axes $R,z,\phi$, and constant anisotropy, so that $\langle v^2_R\rangle\propto\langle v^2_z\rangle$. The equations {were} solved against the observed $v_{rms,*}$ using a Multi Gaussian Expansion of the total mass density profile. 

In the analysis of \citep{vHoudt21} the density profile was parametrized as a sum of two sets of Gaussians: one for the stellar density profile and one for the dark matter profile. They used the single S\'ersic fits from GALFIT to the F814W images to generate the Gaussian expansion of the stellar light. Multi-wavelength information with sufficient spatial resolution to quantify any $M/L$-gradients is currently not available for the majority of our sample, so they assumed a constant $M/L$ within each galaxy. For the dark matter component they assumed a spherically symmetric NFW-profile \citep{Navarro96} augmented with the relation between halo mass, concentration and redshift from \citet{Dutton14} to approximate the concentration parameter. {We note that the inclusion of a dark matter component serves to account for any deviation from the assumption that light follows mass. In particular, stellar $M/L$-gradients can be represented by a (negative or positive) dark matter component.}

The model has five free parameters: $M/L$, the virial mass of the dark matter halo $M_{200}$, the inclination $i$, the anisotropy parameter $\beta_z\equiv 1-\langle v^2_z\rangle/\langle v^2_R\rangle$, and the slit centering $x_0$. The models were convolved with a Moffat kernel with an effective seeing that was determined for each source individually{ as follows: t}he kernels were determined by comparing the radial light profiles of the spectra, i.e., the part of the galaxy covered by the slit, with models obtained by degrading the high-resolution F814W images and imposing a virtual slit for a range of seeing conditions. If the kernel shape was badly constrained, they set the parameter describing the wings of the Moffat kernel to $4.765$ \citep{Trujillo01}. For each galaxy, the model was optimized using a Bayesian approach, specifically the MCMC implementation of \citet{ForemanMackey13} to sample the {posterior probability distribution}.

While they found degeneracies between stellar mass and $M_{200}$, the total dynamical mass was in general well constrained. The best-fit models moreover have a median reduced $\chi^2=1.03$.

\subsection{Emission line dynamical masses}\label{sec:method_gas}

\subsubsection{Data cleaning}

\begin{figure*}
\centering
\includegraphics[width=0.99\textwidth]{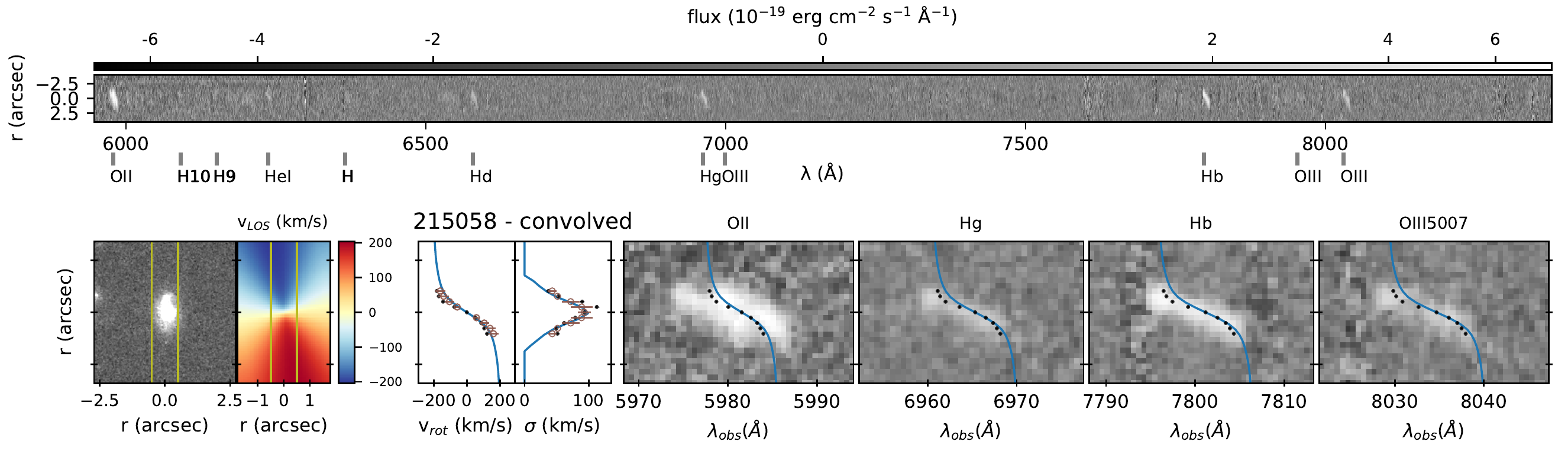}
\caption{\label{fig:examples} Example fit of ID:215058. Top: continuum subtracted spectrum with emission lines{ on an arcsinh scale}. Bottom (from left to right): {HST	}/ACS/F814W image with slit (vertical bars); the model LOS velocity field integrated along the line of sight; the pPXF rotation curve (black data points) and symmetrized curve (open symbols) with best fit model in blue; the pPXF dispersion profile; a zoom-in on the brightest individual emission lines with the pPXF rotation curves and best-fit models. When present, red crosses signify outliers removed from the analysis.}
\end{figure*}

\renewcommand{\thefigure}{\arabic{figure} (Cont.)}
\addtocounter{figure}{-1}

\begin{figure*}
\centering
\includegraphics[width=0.99\textwidth]{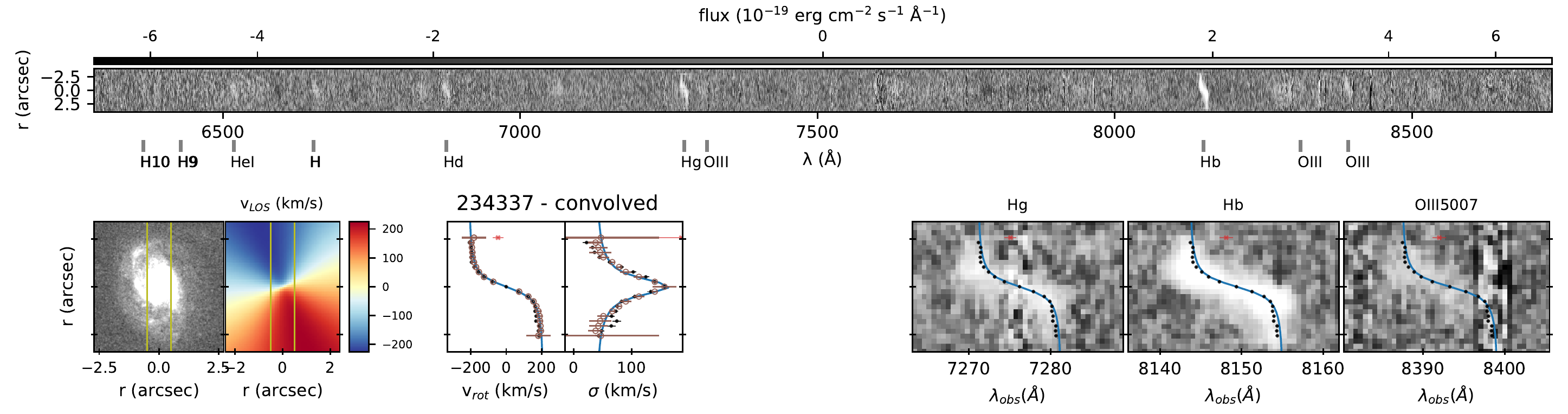}
\caption{Example fit of ID:234337.}
\end{figure*}

\addtocounter{figure}{-1}

\begin{figure*}
\centering
\includegraphics[width=0.99\textwidth]{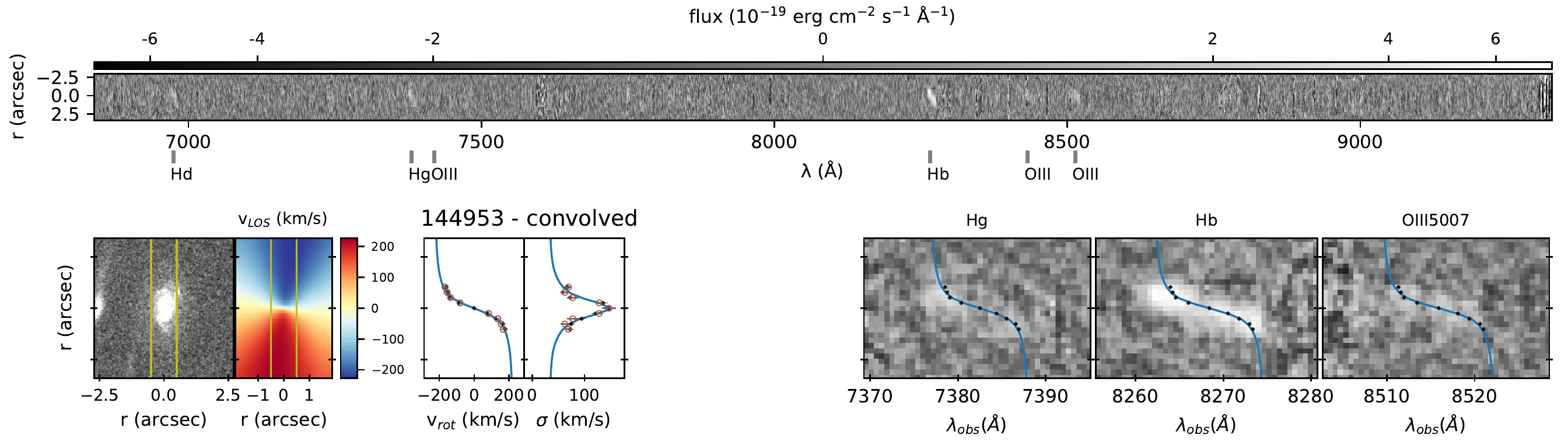}
\caption{Example fit of ID:144953.}
\end{figure*}

The ``emission line dynamical masses'' are the dynamical mass estimates based on the warm ionized gas component of the spectra\footnote{We note that not all astronomical emission lines originate from ionized gas, e.g., in the case of CO and HI, but these are not the target of our current study.}. {For consistency with the stellar dynamics, } we symmetrized the data, {pinpointing and }excluding outlying datapoints. We assumed the kinematic centers of the galaxies were the same as their photometric centers, i.e., at pixel 40. For each symmetrized datapoint at radius $r_i$ from the photometric/kinematic center, we ideally had two datapoints from the rotation curve. If such a pair was available, we calculated the weighted mean and its corresponding uncertainty. If the difference between the pair exceeded the weighted uncertainty, we replaced the latter by this difference. This was often the case, as the pPXF uncertainties are generally very small. If however only one datapoint from only one side of the rotation curve was available, we kept it, but added the median of the differences of all available pairs in quadrature to its pPXF uncertainty. In this way parts of the symmetrized rotation curves that were calculated from only one datapoint do not appear better constrained, i.e., with smaller uncertainties than those based on pairs. {Outliers were flagged after fitting an arctangent or Gaussian to the symmetrized rotation curve or dispersion profile. Datapoints were considered outliers if the rotational velocity deviated at $20\sigma$ significance from the arctangent, if the dispersion deviated at $5\sigma$ from the Gaussian or if a datapoint deviated at the $4\sigma$ level for both rotation and dispersion at the same time. The process, including symmetrization, was repeated iteratively, flagging only one potential outlier at a time, starting with those furthest from the kinematic center, preventing strong influence from potential outliers on the arctangent/Gaussian fits. Furthermore, an outlier was only flagged if it fulfilled the requirements also for fiducial kinematic centers at pixel rows 39 and 41. An example of an outlier can be seen in Figure \ref{fig:examples} for the galaxy with source ID 234337. There is a clear outlier from the rotation curve at a distance of $\sim2\arcsec$ from the center (indicated with a red cross).}

In Figure \ref{fig:examples} we show, for the ionized gas, both the ``raw'', i.e., pPXF generated, datapoints (bullets) and corresponding uncertainties, as well as the symmetrized data (open circles). We note that as the kinematic center is fixed at $r,v_{rot}=(0,0)\ \mathrm{kpc,\ km\ s^{-1}}$, the middle datapoint of the rotation curve has effectively been removed. 

{We note that in some cases the kinematic center does not appear to lie exactly at pixel 40. E.g, the galaxy with ID 215058 (top panel of Figure \ref{fig:fig1}) exhibits a slight asymmetry between fit and data. We have tried to run fits (see Section \ref{sec:HELA} below) with the kinematic center as a free parameter, but it was usually hard to establish whether those results represented the ground truth. In this particular case, the data is affected by the skyline close to the \hb\ line. In other cases, the light profile could be asymmetric, leading to a false detection of an offset from pixel 40. In general, we found that leaving the kinematic center free resulted in better fits for some sources, but worse for others. We concluded that we do not have enough information in our data to independently constrain the kinematic center. }

\subsubsection{Extracting dynamics from slit spectra}\label{sec:HELA}

Our custom code {is based on similar principles as other state of the art algorithms currently used for large surveys, using forward modeling of a 3D disk instead of measuring velocity offsets directly from the sheared emission line{, which leads to underestimated rotational velocities, depending on the width of the slit and $|\Delta\mathrm{PA}|$} \citep[for the latter method see, e.g.,][]{Miller11,Straatman17}.} {It is most similar to }the work of \citet{Price16} and \citet{Price20} who developed the MisFIT algorithm. Other methods that build from a 3D representation of a galaxy are DYSMAL \citep{Cresci09,Davies11}, $^{3D}$BAROLO \citep[especially for higher resolution and/or IFU data;][]{DiTeodoro15} and GBKFIT \citep{Bekiaris16}. 

It is important to note that our model is not a simulator of galaxy mass, in the sense that it would enable a decomposition of the contribution of stellar, gas and halo components to the observed rotation curve. %
Another way to describe the functionality %
is that it provides a more accurate way to account for optical effects than, e.g., a direct fit of a convolved arctangent to a rotation curve. The optical effects include beam-smearing, slit misalignment, and the mixing of light from different velocity regions captured by wide slits (e.g., $1\arcsec$ for LEGA-C). A simple fit to a rotation curve with a $\sin{i}$ factor to correct for inclination and $\cos{\mathrm{|PA|}}$ factor to correct for slit-misalignment will underestimate the rotational velocity, by an amount depending on the width of the slit and the size of the galaxies \citep{DiTeodoro15,Straatman17}.

Our code generates a model of a gas disk and a simulated observation with a telescope instrument. The physical properties of the disk (the luminosity profile, the steepness of the rotation curve, and whether the velocity dispersion is uniform) are parametrized with mathematical prescriptions and folded into a data cube. The cube is rotated according to inclination and position angle, integrated along the line-of-sight, and convolved with a PSF, resulting in a velocity and dispersion map. These can be reduced to an emission line by imposing a virtual slit, and further reduced into a rotation curve and dispersion profile. 

For this work, we assumed an exponential light profile for the ionized gas in each galaxy. {This was motivated as e.g., \citet{Pizzella04} found that the dynamics of gas and stars in nearby galaxies are consistent with the gas residing in a rotationally {supported }disk.} Inclinations were derived from the axis ratios ($q$) as:

\begin{equation}
\label{eq:q0}
\cos{i}=\sqrt{\frac{q^2-q_0^2}{1-q_0^2}}
\end{equation}
{Following convention we used $q_0=0.19$ \citep{Pizagno07}, with $q_0\simeq0.1-0.2 $ representing the intrinsic flattening of spiral galaxies \citep{Haynes84}.} Galaxies with $q<0.19$ were assumed to be maximally inclined. The scalelengths of the disks, $R_s$, were determined from the GALFIT fits to the F814W images -- we used $R_s=R_e/1.678$, with $R_e$ the half-light radius along the semimajor axis. This is the conversion from half-light radius to exponential disk radius in the case of an ideal, infinitely thin disk. We additionally assumed that each disk has an exponential profile in the z-direction, with a scalelength given by $h_z=0.19\ R_s$. We tested the effect of systematically changing $R_s$ to a larger or smaller value than implied by the stellar light or leaving $R_s$ free in the fit. This did not systematically change our results. {Similarly, we tested implementing the S\'ersic profile of the stars, but found no significant changes. In general,} changes to the light profile are subdominant in the model and act on the few percent level. This is due to the parametric nature of our method. The light profile only acts as a weight on $v_{rot}(r)$ and $\sigma_0$, but does not change the dynamical parameters as would be the case for a model with different mass components. 

The rotation curves were parametrized as:

\begin{equation}
v_{rot}(r)=\frac{2}{\pi}V_{t} \arctan{\frac{r}{R_t}}
\end{equation}
with $V_t$ the tangential, maximum velocity and $R_t$ a scaling factor that determines the steepness of the curve. The intrinsic dispersion was parametrized as a constant and isotropic $\sigma_0$, such that the LOS component amounts to $\sigma_0/\sqrt{3}$, irrespective of inclination. The spatial center was not free in the fit, but was set to correspond to the peak brightness of the continuum when we symmetrized the rotation curves. 

The code assumes a smooth, symmetric light distribution, neglecting potentially clumpy structures or the effects of dust obscuration. It is likely that our results will be affected by this simplification. Recently, \citet{Varidel19} found lower intrinsic dispersions using a flexible code that corrects for clumpy light profiles for $z\sim0$ SAMI galaxies compared to models assuming smooth distributions. Our data at $0.6\leq z<1$ does not have enough resolution to model the true distribution of the gas. A detailed investigation of dust obscuration is also not within the scope of this work as it would reach into the realm of hydrodynamic simulations. Overall, the simplifications underlying the model were tailored to the information in the data{ and consistent with other works in literature, thus allowing a clear comparison of their effect against the stellar dynamics}.

The absolute luminosity of the model disk was not taken into account, as we fit only the rotation curves and the dispersion profiles. Alternatively, we could have chosen to fit the emission lines directly. However, in most cases the observed light profile is not {spatially} symmetric, which makes it difficult to determine the kinematic and spatial center. To account for differences in brightness, one can try to match the model light profile to the data on a row-to-row basis after the model emission line has been generated {as was done by} \citet{Price16}, but this is a similar level of post-processing as reducing the model emission line to a rotation curve and dispersion profile. Moreover, individual emission lines are often contaminated by skylines, which can be compensated for by taking into account a set of multiple emission lines at the same time, so we prefer to fit directly to the pPXF {rotation curves and dispersion profiles}.

We used PyMultinest \citep{Buchner14}  to sample the parameter space of the three free parameters: $V_t$, $\sigma_0$, and $R_t$. We set flat priors for $V_t$ and $\sigma_0$, with $V_t\in[-600,600]\ \mathrm{km\ s^{-1}}$, $\sigma_0\in[5,400]\ \mathrm{km\ s^{-1}}$. {To counter the effect of beam smearing, w}e set an exponential prior for $R_t$ with $R_t\in[0.05,2.2R_s]$, skewed to give high probabilities for small values{ (steep slopes)}. To convolve the LOS ($x,y,\lambda$)$-$cubes, we used the same Moffat kernels as used for the JAM modeling described above.

\renewcommand{\thefigure}{\arabic{figure}}

\begin{table*}
\centering
\caption{\label{tab:results} Emission line dynamic quantities. This Table is published in its entirety in the machine-readable format. A portion is shown here for guidance regarding its form and content.\\$^{a,b}$: limited spatial extent for either gas or stars (radial extent $<1.2\times R_e$ (a) or $<1.2\times 2R_e$ (b)). }
\small
\begin{tabular}{l | c c c c c c}
\multicolumn{7}{c}{main sample (\ns$\leq2.5$)} \\
\hline
ID & radial extent ($\arcsec$) & \vrot\ ($\mathrm{km\ s^{-1}}$) &  $v_{rot,2R_e}$ ($\mathrm{km\ s^{-1}}$) & \sig ($\mathrm{km\ s^{-1}}$) & \mgas$_{R_e}$  & \mgas$_{2R_e}$ \\
\hline
27426 & 0.82 & 215.2 $\pm$ 16.5 & 225.3$^{b}$ $\pm$ 19.4 & 59.2 $\pm$ 15.9 & 10.74 $\pm$ 0.12 & 11.11$^{b}$ $\pm$ 0.17 \\
28012 & 0.615 & 192.1 $\pm$ 11.1 & 203.0$^{b}$ $\pm$ 14.1 & 51.0 $\pm$ 8.3 & 10.5 $\pm$ 0.09 & 10.87$^{b}$ $\pm$ 0.11 \\
28512 & 1.845 & 231.7 $\pm$ 25.9 & 247.1$^{b}$ $\pm$ 35.8 & 78.3 $\pm$ 9.3 & 11.03 $\pm$ 0.15 & 11.43$^{b}$ $\pm$ 0.23 \\
28639 & 0.82 & 87.1 $\pm$ 26.0 & 94.5$^{b}$ $\pm$ 30.7 & 114.3 $\pm$ 7.4 & 10.39 $\pm$ 0.13 & 10.93$^{b}$ $\pm$ 0.13 \\
28706 & 0.615 & 253.6 $\pm$ 19.0 & 286.4$^{b}$ $\pm$ 33.8 & 58.0 $\pm$ 28.1 & 10.73 $\pm$ 0.09 & 11.15$^{b}$ $\pm$ 0.15 \\
(...)\\
\hline
\\
\multicolumn{7}{c}{high \ns\ star forming galaxies} \\
\hline
ID & \vrot\ ($\mathrm{km\ s^{-1}}$) &  $v_{rot,2R_e}$ ($\mathrm{km\ s^{-1}}$) & \sig ($\mathrm{km\ s^{-1}}$) & \mgas$_{R_e}$  & \mgas$_{2R_e}$ \\
\hline
10462 & 0.615 & 89.8 $\pm$ 4.9 & 97.5$^{b}$ $\pm$ 9.2 & 109.1 $\pm$ 3.0 & 10.28 $\pm$ 0.06 & 10.81$^{b}$ $\pm$ 0.08 \\
10902 & 0.82 & 293.9$^{a}$ $\pm$ 10.5 & 298.2$^{b}$ $\pm$ 12.3 & 54.9 $\pm$ 4.7 & 11.4$^{a}$ $\pm$ 0.07 & 11.73$^{b}$ $\pm$ 0.09 \\
28945 & 0.615 & 394.4 $\pm$ 7.5 & 412.8$^{b}$ $\pm$ 7.8 & 87.1 $\pm$ 9.1 & 11.08 $\pm$ 0.06 & 11.44$^{b}$ $\pm$ 0.06 \\
28948 & 0.615 & 226.5$^{a}$ $\pm$ 6.0 & 231.1$^{b}$ $\pm$ 6.9 & 89.0 $\pm$ 8.1 & 11.05$^{a}$ $\pm$ 0.06 & 11.42$^{b}$ $\pm$ 0.08 \\
32934 & 1.23 & 241.4$^{a}$ $\pm$ 16.8 & 247.2$^{b}$ $\pm$ 21.3 & 113.4 $\pm$ 15.4 & 11.34$^{a}$ $\pm$ 0.11 & 11.74$^{b}$ $\pm$ 0.16 \\
(...)\\
\hline
\\
\multicolumn{7}{c}{low \ns\ quiescent galaxies} \\
\hline
ID & \vrot\ ($\mathrm{km\ s^{-1}}$) &  $v_{rot,2R_e}$ ($\mathrm{km\ s^{-1}}$) & \sig ($\mathrm{km\ s^{-1}}$) & \mgas$_{R_e}$  & \mgas$_{2R_e}$ \\
\hline
110805 & 0.82 & 104.7 $\pm$ 30.4 & 118.8 $\pm$ 35.1 & 252.2 $\pm$ 15.1 & 10.82 $\pm$ 0.08 & 11.4 $\pm$ 0.1 \\
145923 & 1.23 & 118.1 $\pm$ 25.0 & 126.8$^{b}$ $\pm$ 25.9 & 35.8 $\pm$ 25.5 & 10.41 $\pm$ 0.21 & 10.81$^{b}$ $\pm$ 0.21 \\
200059 & 1.23 & 195.2 $\pm$ 32.7 & 219.9$^{b}$ $\pm$ 53.2 & 236.6 $\pm$ 8.8 & 11.39 $\pm$ 0.11 & 11.92$^{b}$ $\pm$ 0.15 \\
215835 & 1.025 & 188.5 $\pm$ 26.0 & 203.8$^{b}$ $\pm$ 34.6 & 271.8 $\pm$ 8.6 & 11.2 $\pm$ 0.07 & 11.75$^{b}$ $\pm$ 0.09 \\
\hline
\\
\multicolumn{7}{c}{high \ns\ quiescent galaxies} \\
\hline
ID & \vrot\ ($\mathrm{km\ s^{-1}}$) &  $v_{rot,2R_e}$ ($\mathrm{km\ s^{-1}}$) & \sig ($\mathrm{km\ s^{-1}}$) & \mgas$_{R_e}$  & \mgas$_{2R_e}$ \\
\hline
87345 & 1.23 & 68.1 $\pm$ 9.8 & 72.9 $\pm$ 11.7 & 343.4 $\pm$ 4.5 & 11.16 $\pm$ 0.05 & 11.75 $\pm$ 0.06 \\
167044 & 0.615 & 223.1 $\pm$ 153.4 & 245.9$^{b}$ $\pm$ 170.7 & 317.4 $\pm$ 75.4 & 11.1 $\pm$ 0.2 & 11.65$^{b}$ $\pm$ 0.26 \\
209377 & 1.845 & 75.5 $\pm$ 12.8 & 78.6 $\pm$ 13.0 & 251.3 $\pm$ 4.2 & 10.96 $\pm$ 0.05 & 11.55 $\pm$ 0.06 \\
225097 & 0.615 & 230.4 $\pm$ 18.5 & 289.4 $\pm$ 40.6 & 120.4 $\pm$ 32.2 & 10.6 $\pm$ 0.12 & 11.13 $\pm$ 0.17 \\
237641 & 1.23 & 485.3 $\pm$ 34.4 & 506.3$^{b}$ $\pm$ 40.3 & 207.2 $\pm$ 17.8 & 11.69 $\pm$ 0.09 & 12.08$^{b}$ $\pm$ 0.11 \\
\hline
\end{tabular}
\end{table*}

Examples of {convolved data and} fits are shown in Figure \ref{fig:examples}. A more detailed description can be found in Appendix \ref{app:hela}. {The dynamical results (\vrot, \sig) are shown in Table \ref{tab:results}. We defer a future analysis of \vrot\ and \sig\ to a future paper. However we verified as a sanity check that our range of \sig\ values and \vc\ values (see Section \ref{sec:hela_results} below) is in agreement with other literature values at $0.6\leq z<1.0$ \citep[e.g.,][]{Ubler17,Ubler19}.}

\subsubsection{Enclosed mass}\label{sec:hela_results}

{To calculate the emission line dynamical masses, we will apply an asymmetric drift correction to the rotational velocity as derived above. We start by assuming hydrostatic equilibrium, so we can write} %

\begin{equation}
\label{eq:hystat}
\frac{v_{rot}(r)^2}{r}=f_g(r)+\frac{1}{\rho(r)}\frac{dp}{dr}
\end{equation}
with $\rho(r)$ the gas density and $f_g(r)$ the value of the gravitational force. $p$ is a pressure term that consists of a turbulent and a thermal part, i.e., $p=\rho(r)(\sigma_r(r)^2+c_s^2)$, with $\sigma_r(r)$ the radial projection of the velocity dispersion of the gas and $c_s$ the sound speed. The zero pressure rotation curve can be obtained by setting $dp/dr=0$ and $V_c(r)^2\equiv f_g\times r$. Thus we can reduce Equation \ref{eq:hystat} to

\begin{equation}
\label{eq:psupp}
V_c(r)^2=v_{rot}(r)^2 - \frac{1}{\rho(r)}\frac{d\rho(r)\sigma_r(r)^2}{d\ln r}
\end{equation}
Here we have ignored thermal pressure similar to \citet{Burkert10}, as the sound speed is in general much lower than the turbulent velocity. The second term in Equation \ref{eq:psupp} represents the asymmetric drift correction.

For a self-gravitating disk with an exponential mass profile and with constant and isotropic $\sigma_r(r)=\sigma_r$, independent of the height above the disk, Equation \ref{eq:psupp} can be simplified to 

\begin{equation}\label{eq:vc}
V_c(r)^2=v_{rot}(r)^2 + 2\frac{r}{R_s}\sigma_r^2
\end{equation}
\citep[][]{Spitzer42,Binney08,Burkert10}. We now assume that we can replace $\sigma_r=\sigma_0/\sqrt{3}$, taking into account the $\sqrt{3}$ factor used above to correct for the LOS projection of the gas dispersion assuming isotropy. The dynamical mass within a radius $r$ can then be written as

\begin{equation}
\label{eq:mdyn2}
M_{dyn,eml}(<r)=\beta\frac{V_c(r)^2r}{G}\\
=\beta\frac{(v_{rot}(r)^2 + k\sigma_0^2/3)r}{G}
\end{equation}
with $\beta=1$ for a spherical system and $k=2\frac{r}{R_s}$.  {Results are shown in Table \ref{tab:results}.}

We note that this combination of $\beta$ and $k$ in Equation \ref{eq:mdyn2} is misplaced, because a self-gravitating disk is not spherical. However, we show in Section \ref{sec:caveats} that at the resolution of our data, geometrical considerations are subdominant for our estimation of dynamical mass. We also note that at very large radii, {the second term of Equation \ref{eq:mdyn2} becomes dominant. This is likely to be non-physical, as the radial dependence of the second term was in this case derived with the density profile of the tracer (ionized gas) in mind, and not the true density profile including {stars and }dark matter}.

\subsubsection{Uncertainties}\label{sec:eml_uncertainties}

\begin{figure*}
\centering
\includegraphics[width=0.49\textwidth]{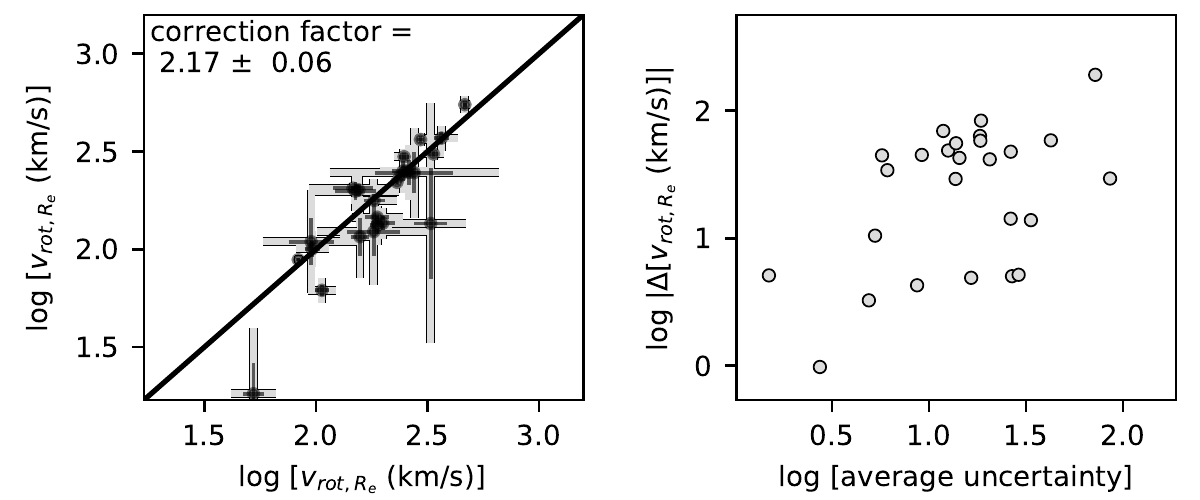}
\includegraphics[width=0.49\textwidth]{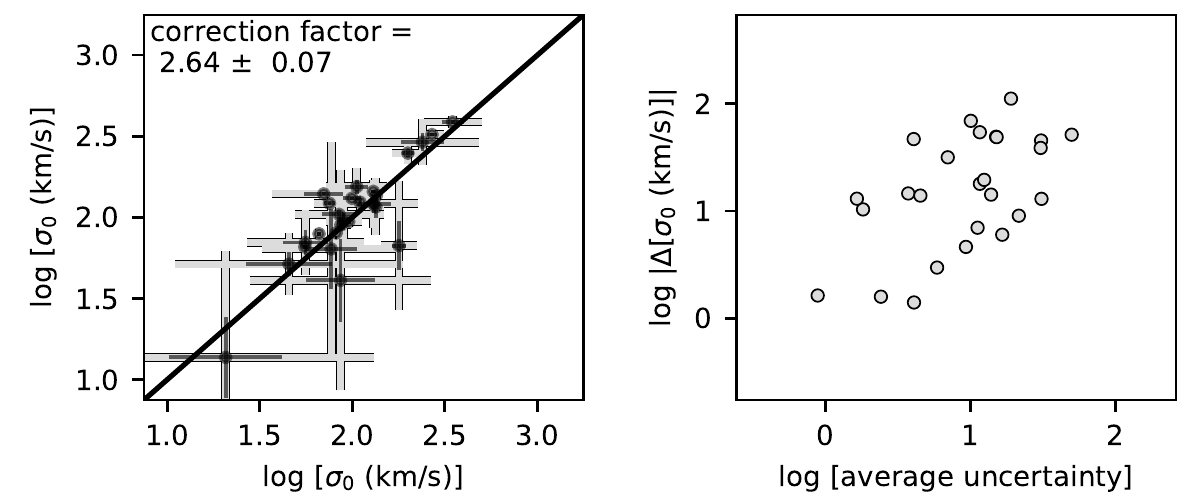}
\includegraphics[width=0.49\textwidth]{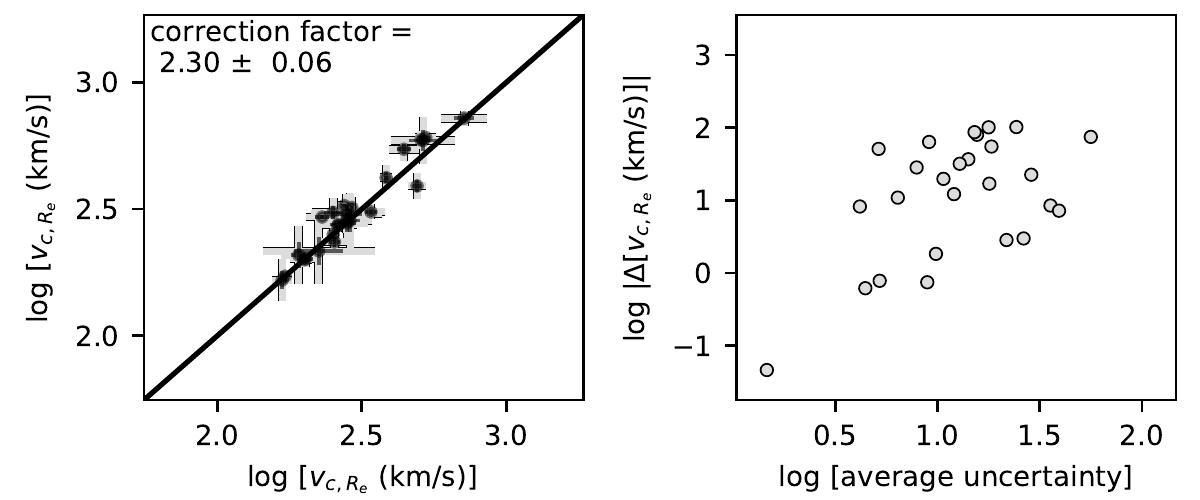}
\includegraphics[width=0.49\textwidth]{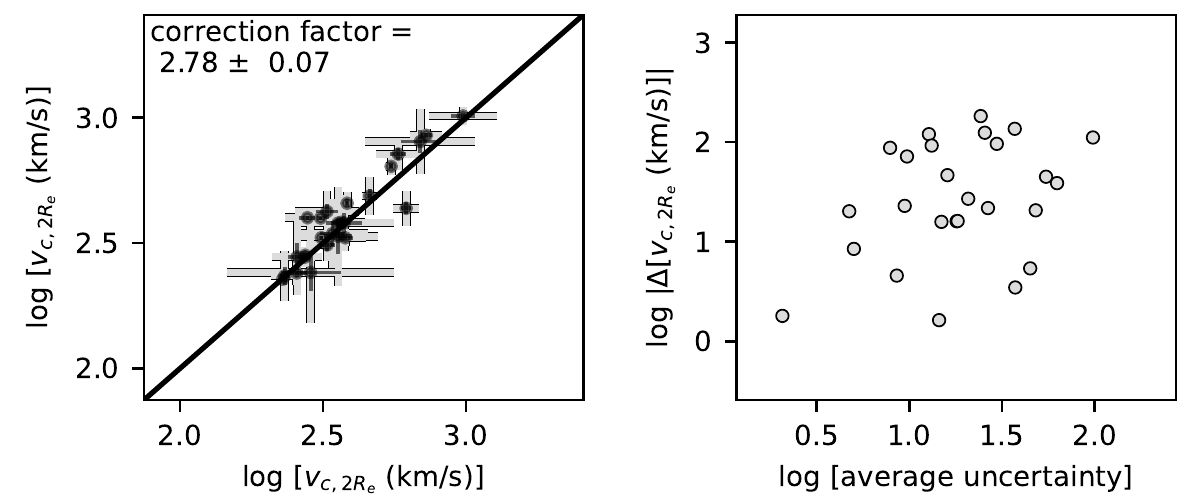}
\caption{\label{fig:dup} Comparison of dynamical measurements of 24 duplicate sources. The uncertainties obtained from the posterior are shown as thin black errorbars in the first of each pair of panels. The absolute differences between duplicate results are proportional to their average uncertainties (second panel of each pair). We normalized the differences by the combined uncertainties and obtained the standard deviation for each variable. The standard deviation is the correction factor needed to make the errorbars consistent with the scatter between duplicates and is printed at the top of each first panel. The corrected uncertainties are shown as thick grey errorbars.}
\end{figure*}

The uncertainties on the dynamic quantities are a combination of the standard deviation of the posterior distribution from the fit, systematics introduced while obtaining the data or during the reduction, and astrophysical assumptions. We performed a test to bring the formal uncertainties from the posterior closer to the actual uncertainties, by comparing duplicate observations of the same sources. The duplicates are observed with different masks, but have similarly long integration times and high $S/N$. Errors in wavelength calibration, acquisition\footnote{Introduced during \textit{observing} by imperfect alignment of the slit onto the source. Note that for the JAM models (Section \ref{sec:jam}) the slit centering was included as the free parameter $x_0$.} and spatial offsets\footnote{Introduced during \textit{reduction} when spectra from different OBs were co-added after determining their spatial centers with a Moffat light profile \citep{Straatman18}.} between OBs will show up as variance between the duplicate results. There are only {seven} duplicate sources in our sample, but in Figure \ref{fig:dup} we show a more extended sample of 26 duplicates. This extended sample has the following additional properties: sources with {\texttt{FLAG\_MORPH}$>0$ or \texttt{SERSIC\_LIMIT}$>0$} are allowed; there are no cut-offs in $n_{S\acute{e}rsic}$, $\mathrm{log\ SFR/M_* (yr^{-1})}$, continuum $S/N$, or \hb\ and \hg\ flux; and non-primary LEGA-C targets are also allowed.
All of these sources should yield the same results between observations, even if the resulting dynamical quantity has no added meaning for this study, and can therefore be used to test systematics in the data.

We found from Figure \ref{fig:dup} that the errors from the posterior distribution, shown in black, are small compared to the variation between duplicate results, although the errors increase with increasing difference between duplicates. For each quantity we take the standard deviation of the differences between duplicate measurements normalized by the combined uncertainties. {If the uncertainties were representative of the true uncertainty, the distribution of these normalized differences would have a standard deviation of one. For under(over)estimated errors, the distribution would have a proportionally larger (smaller) standard deviation. Therefore we can use the obtained standard deviation as a correction factor.} 
The corrected uncertainties are shown in Figure \ref{fig:dup} as thick grey errorbars. We note that the distribution of the differences between duplicates depends on whether a duplicate is assigned to the x-axis or the y-axis. Therefore we randomly shuffled the duplicates between x- and y-axis $1000\times$ and took the median standard deviation as the final correction factor. The correction factors are printed in each panel. %

{The median error after applying the correction on \vc is 19 km/s. This encompasses random errors and systematics in the data, but does not take into account systematics in the model. We discuss this in the next Section. }For the remainder of the paper we apply the correction factor to the uncertainties on the emission line dynamical quantities. 

\section{Comparison of dynamical masses}\label{sec:results}

In this Section we compare the stellar and gas dynamical masses. These are not entirely independent measurements, as they are coupled through the S\'ersic and pPXF fits, although for the pPXF fits the line centroids of the stellar and emission line components are {fit independently}. The key to this study is that the stellar light and emission lines are physically different tracers and that the dynamical models are different in nature and rely on different assumptions. While dynamical masses of distant galaxies have been measured from emission lines in many studies, this is the first time that we are able to validate current methods against stellar dynamical masses.  

\begin{figure*}
\centering
\includegraphics[width=0.8\textwidth]{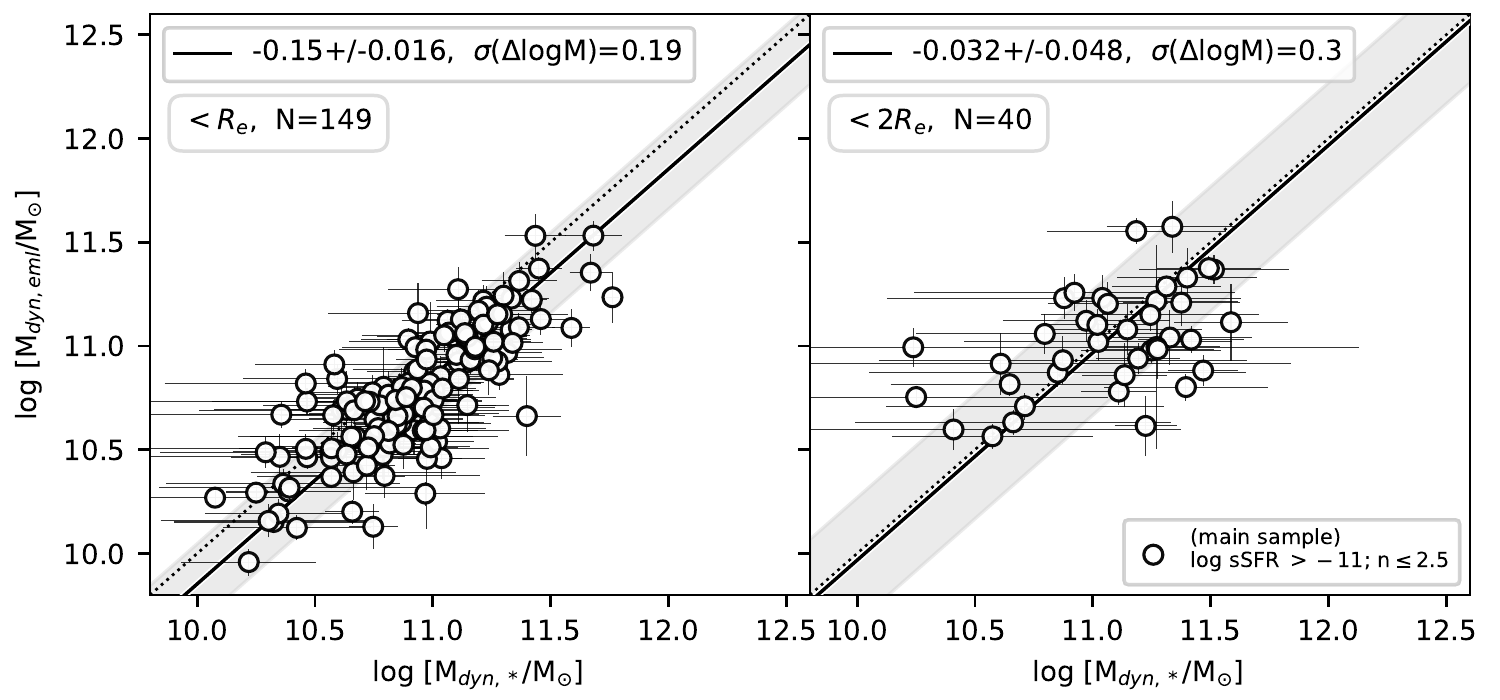}
\includegraphics[width=0.8\textwidth]{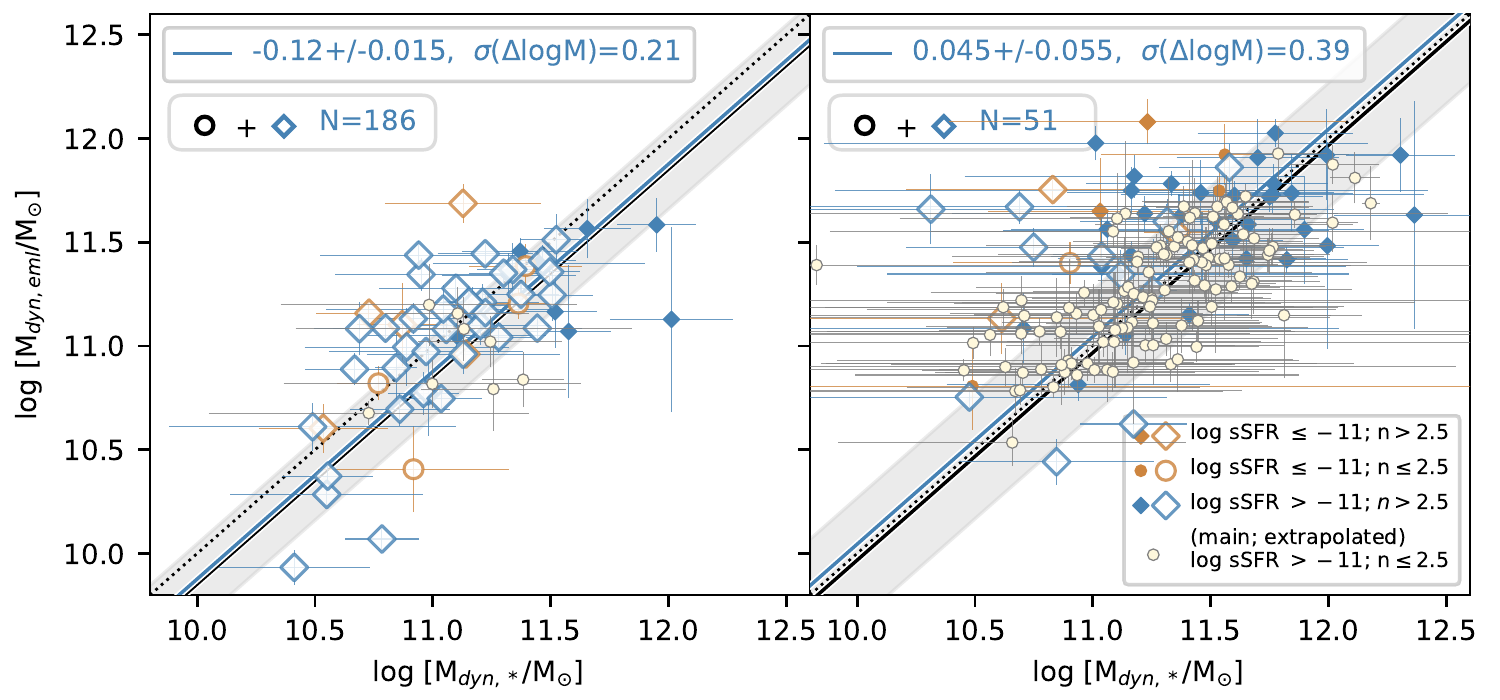}
\caption{\label{fig:fig3} \textit{Top panels}: main result. We show the stellar dynamical mass (x-axis) versus the emission line dynamical mass within $R_e$ (left panel) and $2R_e$ (right panel){ of 149 and 40 {sources with extended rotation curves ($<20\%$ extrapolation of the models)}, respectively}. On average, the emission line dynamical mass is smaller by \mainabsoffset\ dex than the stellar dynamical mass within $R_e$, with a galaxy-to-galaxy scatter of \mainscatter\ dex, which reflects the combined random uncertainty on dynamical mass. %
Within $2R_e$, the gas dynamical masses are in good agreement, but with a larger scatter of 0.3 dex. {\textit{Bottom panels}: sources not included in the main sample. We show high S\'ersic index star forming galaxies {with extended rotation curves }with open blue diamonds, and high and low  S\'ersic index quiescent galaxies with open orange diamonds and {circles}, respectively. {Sources with limited spatial extent ($>20\%$ extrapolation of the models),} including those of the main sample, are shown with smaller, filled symbols. High S\'ersic index star forming galaxies have somewhat higher emission line dynamical masses. However, an analysis of all star forming galaxies {with extended rotation curves} ($N=186$ at $R_e$ and $N=51$ at $2R_e$), results in similar offsets and scatter (blue text {in legend}).}}
\end{figure*}

In the top panel of Figure \ref{fig:fig3} we compare the dynamical masses of the main sample within $R_e$ and $2R_e$. We calculated the median offset $\mathrm{\Delta \left[ log\ M\right]}=\ $\dm\ and average galaxy-to-galaxy scatter $\sigma(\mathrm{\Delta \left[ log\ M\right]})$. We found that on average \mgas\ is \mainabsoffset\ dex lower within $R_e$ than \mjam. We also found an average galaxy-to-galaxy scatter of \mainscatter\ dex. Within $2R_e$ the different mass estimates of the much smaller subsample of sources are in agreement, but with a large formal uncertainty of $12\%$ on the offset and a galaxy-to-galaxy scatter of $0.3$ dex.

It is clear from Figure \ref{fig:fig3} that the formal uncertainties on \mjam\ are larger than on \mgas: a median of 0.23 dex versus 0.075 dex. \citet{vHoudt21} found from a duplicate test that the uncertainties on their mass estimates were on average 50\% larger than the scatter between the duplicates, indicating they might be overestimated due to the large flexibility of the JAM model provided by the inclusion of a dark matter component. However, the duplicate test only takes into account the random component of the uncertainty, but not systematics introduced by e.g., the GALFIT model. They concluded that as the duplicate measurements are not independent, e.g., because the same size was used for duplicate mass estimates, it was best to use a conservative estimate and not apply the duplicate correction. While the dynamical mass measurements here are not entirely independent either, as the same S\'ersic model was used for both, we are in a slightly better position to account for systematics between the different models. The combined random uncertainty on a dynamical mass measurement with LEGA-C is \mainscatter\ dex (see $\sigma(\mathrm{\Delta \left[ log\ M\right]})$ in the top left panel of Figure \ref{fig:fig3}). From the duplicate test of \citet{vHoudt21} it follows that the random component of the uncertainty on the stellar dynamical mass is $0.23/1.5$ dex. Taken together, this leaves $\sqrt{0.19^2-(0.23/1.5)^2-0.075^2}=0.083$ dex of uncertainty due to systematics related to the choice of modeling.

We note that part of the variance in $\mathrm{M_{dyn,*}}$ is due to the broad inclination prior used for the JAM models, irrespective of whether the galaxies are disks or ellipticals. As the galaxies in this study have been selected to have $n_{S\acute{e}rsic}<2.5$ there is a probability that the inclination was underestimated for those seen face-on, leading to underestimated rotational velocities and dynamical masses. For the emission line dynamical masses, on the other hand, we derived inclinations while explicitly assuming the galaxies are disks. Therefore we would expect a number of galaxies with underestimated \mjam\ and overestimated \dm. Indeed we found that 59\% of galaxies with \mjam$<10.5$ have axis-ratio $q>0.8$.

Our main sample is limited to low S\'ersic index galaxies, however there are 47 star forming galaxies with \ns$>2.5$ that pass our selection criteria, of which 37 have extended rotation curves{ reaching $R_e$}. For clarity, all sources that pass the criteria, but do not belong to the main sample, are shown in the bottom panels of Figure \ref{fig:fig3}, including sources {with limited spatial extent}. The high S\'ersic index star forming galaxies (blue diamonds) tend to have somewhat larger emission line dynamical masses{, with an average offset of $-0.026\pm0.040$ dex and scatter of 0.24 dex.}. However, our main result is not strongly biased by the exclusion of these sources: the average offset of all 186 star forming galaxies {with extended rotation curves} is $-0.12\pm0.015$ {, with a scatter of 0.21 dex.}

{There are only 11 high S\'ersic index star forming galaxies with extended rotation curves reaching $2R_e$. For this sample, the average offset is $0.33\pm0.16$ dex with a scatter of 0.53 dex. For the sample of all 51 star forming galaxies with extended rotation curves to within $2R_e$, the average offset is $0.045\pm0.055$ dex with a scatter of $0.39$ dex, again similar to the main sample.}

{It has to be noted that the measured dynamical quantities suffer less from seeing effects at $2R_e$ than at $R_e$. The agreement between the estimates of the dynamical masses within $2R_e$ is, however, likely accidental and due to the adopted uniform, dispersion profile. \sig\ is assumed to be the same at radii of $R_e$ and $2R_e$, but is more heavily weighted for emission line dynamical mass estimates evaluated within $2R_e$ (Equation \ref{eq:mdyn2}). This effect is stronger than the flattening of the rotation curve between $R_e$ and $2R_e$ and leads to relatively larger emission line dynamical masses with increasing radius. It explains the large positive offset for the 11 high S\'ersic index star forming galaxies with extended rotation curves reaching $2R_e$, as these galaxies tend to have larger \sig. The large scatter between the dynamical mass estimates within $2R_e$ (0.3 dex, compared to 0.19 for $R_e$) also implies that the lack of an offset does not necessarily reflect a better agreement. Data of higher spatial resolution are needed to better understand the dynamics of both gas and stars within $R_e$ and $2R_e$.}

T{aken together, t}he small, negative offset of the average \dm\ within $R_e$ and {the negligible difference within $2R_e$ indicate a good correspondence between the emission line and stellar dynamical masses. This finding has direct implications for mass studies of the distant universe: we now have a better understanding of how biased emission line dynamical masses likely are in the absence of expensive continuum observations. For a discussion on the accuracy of the stellar dynamical masses, see Section \ref{sec:caveats}.} 
{The similarity between the different dynamical mass estimates, }where for the emission line dynamical mass we assumed the gas is in hydrostatic equilibrium, suggest there is little room for {strong deviations from equilibrium} dominating the galaxy-wide velocity dispersion. The main conclusion of this paper is {therefore }that we do not see strong evidence for increased linewidths due to non-gravitational motions. 
{In the following Section we will look for residual trends between the mass offsets and other galaxy properties (Section \ref{sec:residual_trends}), compare with dynamical masses that can be obtained without modeling from the dispersion of spatially integrated emission lines (Section \ref{sec:virial_masses}), and discuss potential caveats on our derivation of both the emission line and stellar dynamical masses (Section \ref{sec:caveats}).}

\section{Discussion}\label{sec:discussion}

\subsection{residual trends with other galaxy properties}\label{sec:residual_trends}

\begin{figure*}
\centering
\includegraphics[width=0.99\textwidth]{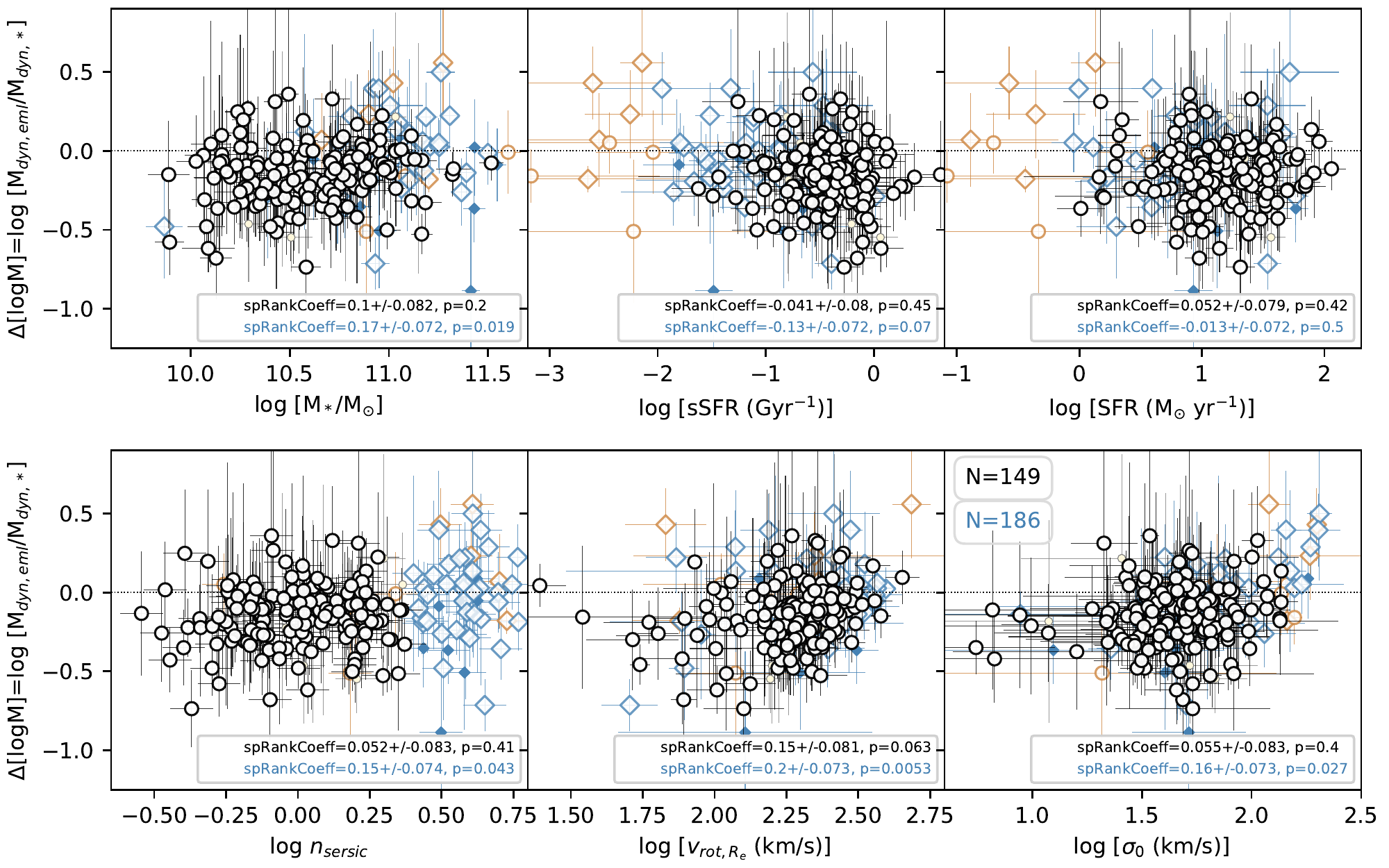}
\caption{\label{fig:fig7} {The ratio between the two dynamical mass estimates (\dm){ within $R_e$} as a function of (from left to right and top to bottom) stellar mass, sSFR, SFR, S\'ersic index, \vrot, and \sig. The symbols are the same as in Figure \ref{fig:fig3}. We found no significant correlations for galaxies in the main sample, suggesting no strong deviations from hydrostatic equilibrium (see main text). When including high S\'ersic index star forming galaxies, we found correlations (blue text in legends) with \ns, \vrot, and \sig. Including these galaxies additionally results in correlations with stellar mass and sSFR, which is then likely a selection effect.}}
\end{figure*}

In Figure \ref{fig:fig7} we show \dm\ within $R_e$ as a function of {stellar mass, sSFR, SFR, S\'ersic index}, \vrot, and \sig. We calculated the Spearman's rank correlation coefficient (``spRankCoeff'' in Figure \ref{fig:fig7}) for each parameter versus \dm{, for the main {sample of galaxies with extended rotation curves,} as well as the sample including high S\'ersic index star forming galaxies}. {W}e do not find any significant correlations ($p<0.05$) between \dm\ and {these parameters for the galaxies in the main sample. Especially interesting is the lack of correlation with SFR, as outflows are often invoked as an explanation for the higher  average intrinsic dispersions of the ionized gas in high redshift galaxies.} \citet{NFS19} found that {the incidence of }SF-driven outflows correlate{s} with the offset from the SFR-mass relation for star forming galaxies. {If the effect of such outflows dominated the emission line dispersions, we would expect to overestimate the dynamical masses and find a positive correlation of \dm\ with SFR.} \citet{Genzel11} showed that outflows can arise from individual clumps or star forming regions within galaxies {at $z\sim2$}, but that these do not affect the galaxy integrated dispersion. \citet{Bassett14} similarly found that the dispersions of the ionized gas and the stars in two $z\sim0.1$ galaxies are well matched, indicative that the ionized gas turbulence, which would otherwise have resulted in comparatively high dispersions, is not dominated by non-gravitational motions caused by stellar feedback in the form of stellar winds or supernovae. {At higher redshift, \citet{Ubler19} found that the measured velocity dispersions were consistent with being primarily driven by gravitational instabilities, with only a sub-dominant contribution from stellar feedback.}
{In general, the lack of a correlation with sSFR that we see in this study implies that the kinetic energy of the ionized gas represents dynamical equilibrium with the gravitational potential.}

In addition to the variables shown in Figure \ref{fig:fig7}, we investigated potential correlations with SFR surface density, $R_e$, redshift, continuum $S/N$, axis-ratio and position angle. We did not find significant correlations for any of these parameters. {However, if we include the high S\'ersic index star forming galaxies, we found modest correlations with stellar mass, S\'ersic index, \vrot\ and \sig. The higher \ns\ galaxies {in some cases appear to follow the light profile of the stars with luminous central regions with large dispersions and so} tend to have larger \sig{, introducing a correlation that is not necessarily representative of the population}. The correlations with stellar mass and sSFR are likely selection effects, as the higher \ns\ galaxies include the most massive galaxies.

{As a final remark we note that these results are limited to a particular mass ($10^{10}\msun \lesssim M \lesssim 10^{11.5}\msun$) and redshift range ($0.6\leq z < 1$). Lower mass galaxies are on average more dispersion dominated \citep[e.g.,][]{Kassin12,Wisnioski15,Price20} or more susceptible to mass loss through outflows. The incidence of AGN with visible broad line components increases towards $\sim70\%$ at $>10^{11}\msun$ \citep{NFS19}, but the potential effect on dynamical mass is not investigated here. At higher redshift the fraction of irregular galaxies increases, as well as the average velocity dispersion of star forming galaxies \citep[e.g.,][]{Kassin12,Wisnioski15,Ubler19,Price20}.}

\subsection{Emission line ``virial'' masses}\label{sec:virial_masses}

So far we have investigated the validity of dynamical mass estimates that can be obtained if the data is sufficiently deep and spatially {extended}. For fainter sources or shorter integration times, dynamical masses can be calculated from the dispersion of the spatially integrated emission lines: $\sigma'_{g,int}$. Using DR2 of LEGA-C, \citet{Bezanson18b} showed that the integrated dispersions of the ionized gas and the stars at $z\lesssim1$ are in good average agreement, with a 0.13 dex intrinsic scatter. This would suggest a $\sim0.24$ dex uncertainty on dynamical masses based on $\sigma'_{g,int}$. We can now test how well these ``virial'' masses correspond using the spatially {extended} stellar dynamics.

We use the following definition for virial mass:
\begin{equation}
\label{eq:mvir}
M_{vir,eml}=\beta(n)\frac{{\sigma'}_{g,int,corr}^2R_e}{G}
\end{equation}
with $\beta(n)$ the analytical approximation for the virial coefficient by \citep{Cappellari06}:
\begin{equation}
\label{eq:vircoff}
\beta(n)=8.87-0.831n+0.0241n^2
\end{equation}
which we note is originally valid for spherical, isotropic distributions, {and $\sigma'_{g,int,corr}$ the inclination corrected dispersion following van Houdt et al. ({in prep}):}
\begin{equation}
\sigma'_{g,int,corr}=\sigma'_{g,int}\left( 0.87+ 0.39e^{-3.78(1-q)} \right)
\end{equation}
The correction was based on the stellar dynamics while using a more complete sample from LEGA-C including more quiescent galaxies, but there was no clear difference between high and low S\'ersic index galaxies.
Assuming the stellar light reflects most of the mass profile within $R_e$ and accounts for half of the total within that radius, we expect that $\mathrm{M_{dyn,*}}\ (R_e)\simeq 0.5 \mathrm{M_{vir,eml}}$.

\begin{figure*}
\centering
\includegraphics[width=0.8\textwidth]{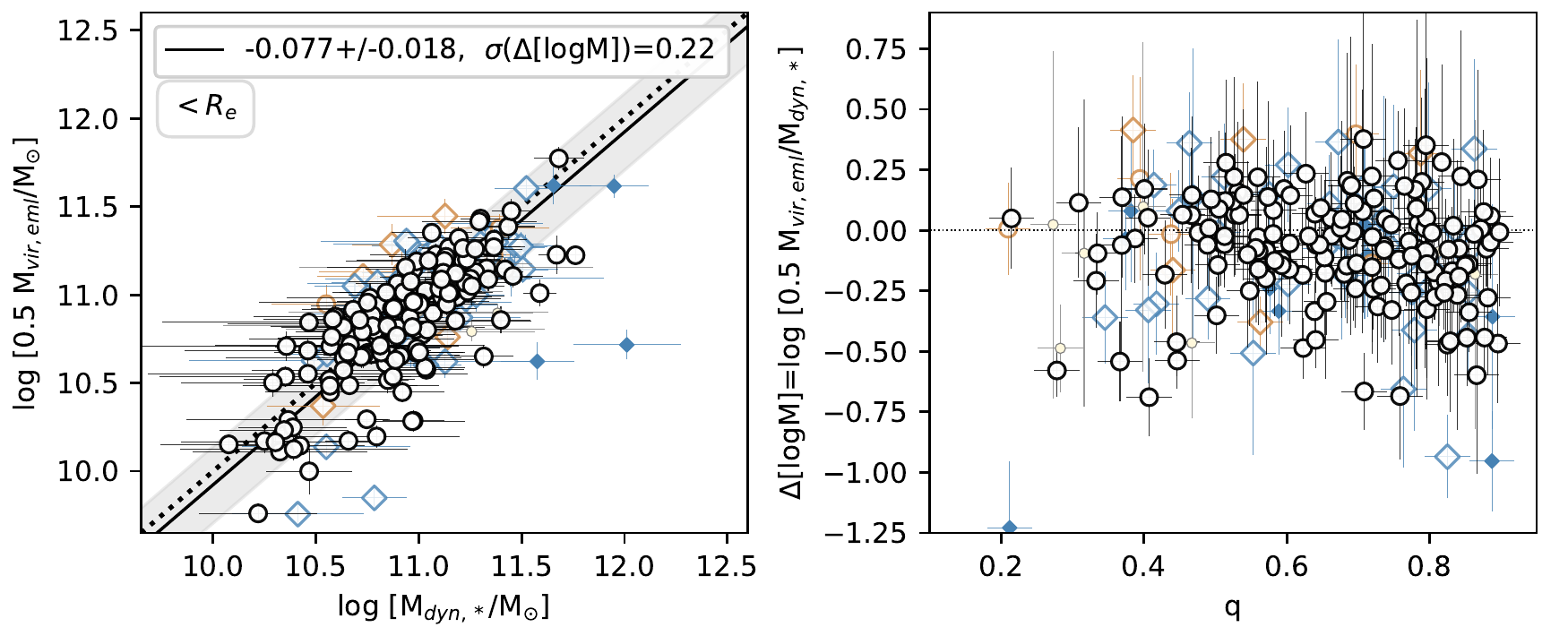}
\includegraphics[width=0.8\textwidth]{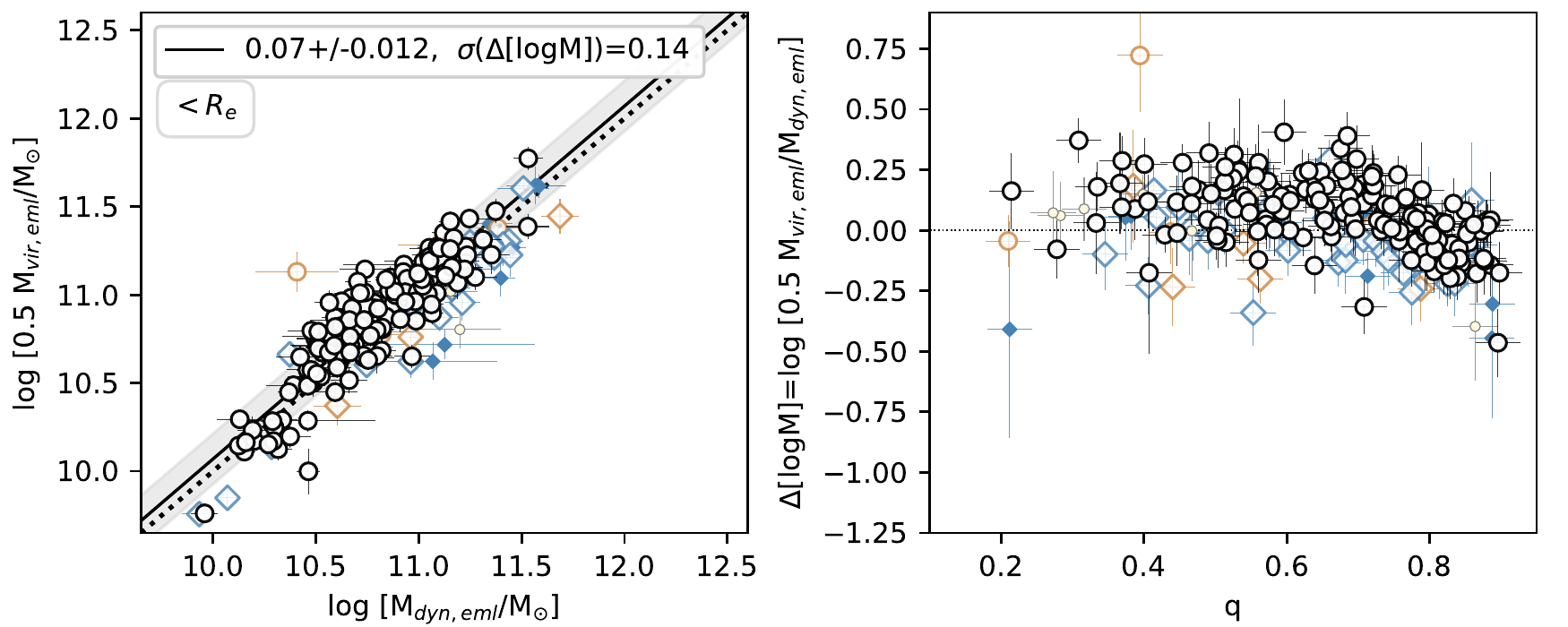}
\caption{\label{fig:virial} \textit{Top left:} a comparison of the ``virial'' masses and the stellar light dynamical masses within $R_e$. The ``virial'' masses are based on the dispersion of the spatially integrated emission lines {corrected for inclination} ($\sigma'_{g,int,corr}$). The low mass outliers are mostly face-on disks. \textit{Top right:}{a clear dependence on axis ratio is visible, but in general} accurate dynamical masses can be obtained from $\sigma'_{g,int.corr}$ if the axis ratio $q<0.8$. \textit{Bottom left:} a comparison of the ``virial'' masses and the emission line dynamical masses. This relation is tighter as the same dynamical tracer was used. \textit{Bottom right: } {dependence of mass offset on axis ratio. The inclination correction on $\sigma'_{g,int}$ was derived based on the stellar dynamics by van Houdt ({in prep}). The residual trend suggests that the ionized gas exhibits a flatter morphology than the stars.}}
\end{figure*}

In the top panels of Figure \ref{fig:virial} we compare the emission line based virial masses to the JAM results. {The discrepancy between the uncertainties on these two quantities is formal: the half-light radius ($R_e$) and integrated dispersion ($\sigma'_{g,int}$) are well determined, even with a duplicate factor included \citep{Straatman18,vdWel21}, whereas the JAM models are marginalized over multiple parameters. }We found the following statistics: an offset of $-0.077\pm0.018$ dex, with a scatter of 0.22 dex. %
{If we adopt \ns$=1$ in Equation \ref{eq:vircoff}, because we did so for the emission line model, we find a similar offset and scatter of $-0.072\pm0.018$ dex and 0.21 dex, respectively.}

There is a tail of galaxies at the low-mass end with $\mathrm{M_{dyn,*}}\ (R_e)>0.5 \mathrm{M_{vir,eml}}$. Upon further inspection {we found a trend with axis ratio (top right panel). The galaxies at the low mass end} constitute the most face-on galaxies in the sample. The median $q$ for galaxies with $0.5 \mathrm{M_{vir,eml}}<10.5$ is $q=0.83$. The ``virial'' masses of these galaxies {could be} underestimated {if} no {correct} inclination correction is applied. %
If we consider only galaxies with $q<0.8$, we found an offset of $-0.056\pm0.020$ dex, with a scatter of 0.21 dex.

{In the bottom panels of Figure \ref{fig:virial} we compare the virial masses to the extended emission line dynamical masses instead of the JAM results. The offset is $0.07\pm0.012$ dex, with a scatter of 0.14 dex ($0.11\pm0.012$ dex and 0.12 dex for $q<0.8$ galaxies). The smaller scatter indicates that $\sigma'_{g,int}$ corresponds better to \vc\ {measured} from the same emission lines than from the JAM models. From the bottom right panel we %
also note a residual trend with $q$. {These trends indicate }that the correction {of van Houdt et al. ({in prep})} is not strong enough for many galaxies with high $q$, suggesting the ionized gas exhibits flatter morphologies than the stars. We tested this assuming thinner disks by varying $q_0$ in Equation \ref{eq:q0}, but found no change in our results. Our data is not sufficiently resolved to test this further. IFU data,{ which has more extended spatial coverage, has better potential to investigate the morphology of the ionized gas. }We note however {the residual dependence }could in part be a selection effect {from not including quiescent galaxies}. We investigated this by reproducing the top right panel of Figure \ref{fig:virial} with the original estimator $\mathrm{M_{vir,*}}$ of van Houdt et al. ({in prep}). This is not shown here, {but we essentially reproduced their Figure 2, except limited to our main sample.} While we found less strong outliers, we did find a bias towards those galaxies with low $\mathrm{M_{vir,*}/M_{dyn,*}}$ at high axis ratio, compared to the full sample of van Houdt et al. ({in prep}) including quiescent galaxies.

\begin{figure}
\centering
\includegraphics[width=0.42\textwidth]{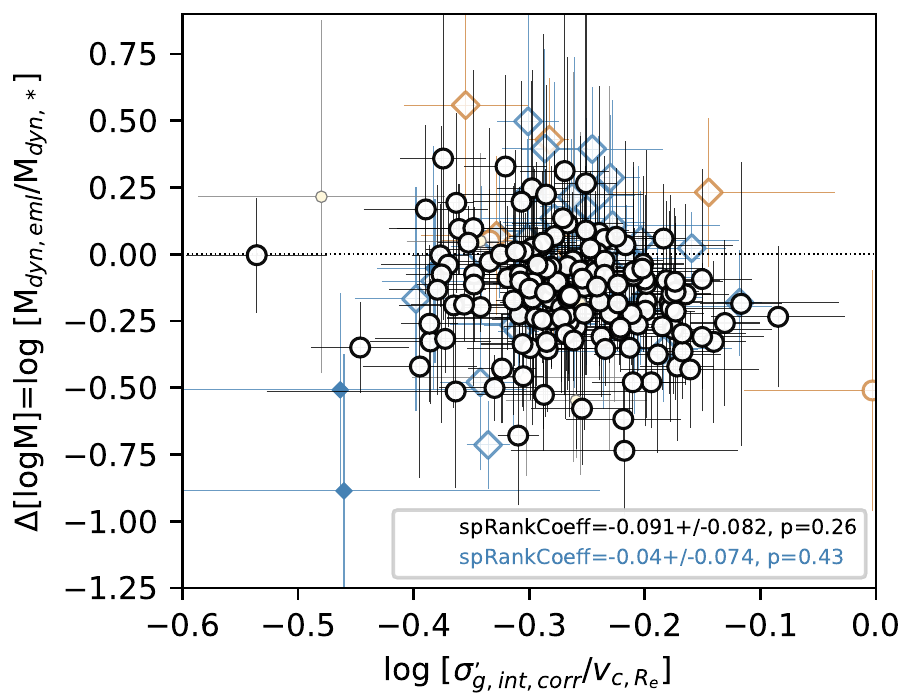}
\caption{\label{fig:fig12} {The ratio between the two dynamical mass estimates (\dm) as a function of $\sigma'_{g,int,corr}/v_{c,R_e}$.  
There appears to be a trend, which is not significant, between \dm\ and $\sigma'_{g,int,corr}/v_{c,R_e}$. The virial masses based on $\sigma'_{g,int,corr}$ show a similar correspondence to \mjam\ as the modeled emission line dynamical masses, so this result may indicate there is space left to improve $v_{c,R_e}$ with a better understanding of pressure support in the gas.}}

\end{figure}

The correspondence between the emission line virial masses and the stellar dynamical masses, especially for $q<0.8$ galaxies, is remarkable, since no extra modeling was done. However, we can turn this fact around and conclude that even though modeling of spatially extended rotation curves might lead to more accurate rotational velocities and intrinsic dispersions, this is negatively balanced by the assumptions that go into Equation \ref{eq:vc}. If there is room for improvement on $v_c$, there should be a trend between $\sigma'_{g,int}/v_{c,R_e}$ and \dm, as deviations of $v_c$ relative to $\sigma'_{g,int}$ should lead to deviations in emission line dynamical mass. In Figure \ref{fig:fig12} we show that there is such a trend, although it is not significant. This trend is not driven by axis ratio, as it is still present after applying the correction of van Houdt et al., ({in prep}). {We note it may be possible that the use of long slit data limits the accuracy of our estimates of \vrot\ and \sig\ compared to what can be achieved IFU data. We verified in Section \ref{sec:HELA} that the range of our dynamical results (e.g., \sig) are not systematically biased compared to other works. We therefore expect that a significant step forward in our understanding can be obtained by using high resolution data to measure e.g., dispersion gradients in perhaps more local galaxies.}

\subsection{Caveats on the models}\label{sec:caveats}

In this study we have assumed that \sig\ is uniform and isotropic. The {analytical} implication is that the vertical scale height of a disk increases exponentially with radius{, because we have presumed that the mass profile follows that of the tracer ionized gas to obtain} Equation \ref{eq:vc}. %
{Disk flaring {has} been found for nearby galaxies \citep[e.g.][]{Bacchini19}, but we note that the flaring implied in Equation \ref{eq:vc} ($h\propto e^{r/R_s}$, with $h$ the scale height of the disk) is probably more severe than reality.} It is also reasonable to assume the situation in most galaxies is more complex, with local or global dispersion gradients. Furthermore, our current prescription for pressure support is based on the assumption of a self-gravitating{, smooth (no clumps),} exponential disk. In reality, the morphology of a galaxy is more complicated. This is illustrated in recent studies of the IllustrisTNG and VELA simulations (\citet{Wellons20} and \citet{Kretschmer21}, respectively). \citet{Kretschmer21}, found that the central and dark matter components of the galaxy system effectively imply lower values of $k$. This would introduce an offset at $2R_e$ and a stronger offset at $R_e$ in Figure \ref{fig:fig3}. 
\citet{Wellons20}{ as well as} \citet{Kretschmer21} {also }found that corrections from anisotropic and non-constant dispersions, as well as from non-spherical potential wells, should be taken into account, but this is simply not possible with the current data. Our current prescription is very similar to those of \citet{Wisnioski18} and \citet{Price20}, who use factors of $\beta=1.09^{-1}$ and $\beta=1.064$, respectively, to account for deviations from the spherical case. 

To further investigate the effect of geometry, we may replace Equation \ref{eq:mdyn2} by one based on a razor thin disk morphology:
\begin{equation}
\label{eq:razor_thin}
M_{dyn,eml}(<r)=\frac{V_c(r)^2R_s}{2y^2GB(y)}\left[1-e^{-r/Rs}\left(1+\frac{r}{Rs} \right)\right]
\end{equation}
with $B(y)=\left[I_0(y)K_0(y)-I_1(y)K_1(y) \right]$, $I_n$ and $K_n$ modified Bessel functions and $y=0.5R/R_s$ \citep{Binney08}. We found that this produces an offset at $R_e$, {slightly more negative, but }consistent with our fiducial result, of \dm$=-0.17\pm0.016$. {This is expected, as the gravitational force is stronger in a disk/ring geometry than for nested spherical shells, resulting in a lower dynamical mass for a given $V_c$. We conclude from this exercise that the correspondence between the two dynamical mass estimates depends more strongly on the assumptions leading to $V_c$ than on the adopted geometry.}

{Our results are in agreement with those of \citet{CrespoGomez21}, who found $\mathrm{M_{dyn,eml}/M_{dyn,*}}=1.0\pm0.3$, comparing ionized gas and stellar kinematics of the inner $1-2$ kpc of local LIRGs, and who also found that $\beta$ would not affect their results within the uncertainties. }

Throughout this work we used the stellar dynamical masses as a benchmark to test whether high redshift emission line based dynamical masses are accurate. There {are} several caveats. In the JAM models, a separate gas component was not included. The invisible mass, gas and dark matter, is represented by the NFW component, whose actual distribution is not constrained by the data. 
The stellar mass component, on the other hand, is comparatively well constrained, except for the assumption of a constant $M/L-$ratio, but there may still be additional uncertainties due to the stellar population models and IMF assumed. More importantly, despite the high continuum $S/N$ of the data, the spatial resolution is limited. Bars and spiral arms are not resolved as they would have been for IFU data of nearby galaxies. 

{Nevertheless, large systematic uncertainties due to model choice appear unlikely. \citet{Leung18} have shown that the stellar \vc\ corresponds to within 10\% to the rotational velocity of molecular gas for $z\sim0$ S0-Sd galaxies, using three different methods, including Jeans modeling and an asymmetric drift correction such as we applied here to the ionized gas. }

{The \mainoffset\ dex difference that we found at $R_e$ will partly be driven by the choice of JAM modeling versus the more simple pressure correction that we applied for the gas. To understand this better, applying a similar dynamical mass comparison to the latest generation of IFU studies (SAMI, ATLAS3D, CALIFA) can shed more light on understanding the high redshift results presented here, and help trace systematics.}

{We opted to do this comparison at $z>0.6$ as gas turbulence appears to play an increasing role in gas dynamics at high redshift. The implicit assumption is that pressure corrections do not play a significant role at $z\sim0$, where dynamical masses and scaling relations, such as the Tully-Fisher relation, are built directly from rotational velocity. E.g., \citet{Pizzella04}, \citet{Bassett14} and \citet{CrespoGomez21} showed that the rotation curves and dispersion profiles for the stellar and ionized gas component in nearby spirals, LIRGs and $z\sim0.1$ galaxies are different, with higher rotational velocities for the gas and higher dispersions for the stars, indicating a potentially subtle effect for asymmetric drift corrections. We defer a comparison of $v_{rot}/\sigma_0$ between the two tracers to future work.}

Finally, even if the stellar and emission line dynamical masses are in agreement, they could still suffer from the same, potentially unknown, systematics, perhaps related to beam smearing effects.

\section{Summary}\label{sec:summary}

In this paper we leveraged the ultra deep LEGA-C survey to calculate and compare the dynamical masses of a sample of \finalsample\ star forming galaxies at $0.6<z<1$ with two different physical tracers: the emission lines from the ionized gas and the continuum and absorption lines from the stars. {Out of the main sample, 5\% and 75\% {have limited spatial extent }beyond $R_e$ and $2R_e$, respectively. We calculated statistics for the {subset of} 149 and 40 sources {with extended rotation curves within each radius, respectively}.}

The main result is that emission line based dynamical masses, derived with recent common methodology, are on average underestimated compared to stellar light based dynamical masses: we derived an offset of \dm$=$\mainoffset\ dex for the enclosed mass within $R_e$ (149 sources). Within $2R_e$ the majority of the mass estimates are based on {sources with limited spatial extent}, but we found good agreement between emission line and stellar dynamical mass: \dm$=-0.032\pm0.048$ dex (40 sources). {Even though we excluded high S\'ersic index (\ns$>2.5$) star forming galaxies from the main sample, we found similar results if those sources were included. Quiescent galaxies are not well represented by our data.}

We hypothesized {in the introduction }that the kinematics of the ionized gas reflect some form of equilibrium, so we could apply an asymmetric drift correction using the solution for a self-gravitating disk in hydrostatic equilibrium with an exponential mass profile. The near correspondence between the stellar and emission line dynamical masses {and lack of residual trends with other galaxy properties }indicates there is indeed little room for increased linewidths due to non-gravitational motions. %

We derived similar offsets for ``virial'' masses based on spatially integrated velocity dispersion ($\sigma'_{g,int}$) with an average offset of $-0.077\pm0.018$ dex and scatter of 0.22 dex, compared to the stellar dynamical masses. {This indicates that at similar resolution to our data, even after modeling spatially extended rotation curves, dynamical mass estimates based on an asymmetric drift correction are still too dependent on assumptions to be significantly more accurate. {This may be relevant for future studies {not specifically aimed at detailed dynamical properties}, e.g., {large shallow surveys} probing the distant universe. Keeping in mind the caveat that the good correspondence between dynamical mass based on $\sigma'_{g,int}$ and stellar dynamical mass is specific for this study and cannot be directly extrapolated to higher redshift, such studies {may} benefit from this result by focussing on measuring $\sigma'_{g,int}$ and saving on integration time.}

\acknowledgements{}
We thank the anonymous referee for their helpful and constructive remarks. Based on observations made with ESO Telescopes at the La Silla Paranal Observatory under programme IDs 194-A.2005 and 1100.A-0949 (The LEGA-C Public Spectroscopy Survey). CMSS acknowledges support from Research Foundation - Flanders (FWO) through Fellowship 12ZC120N. This project has received funding from the European Research Council (ERC) under the European Union’s Horizon 2020 research and innovation programme (grant agreement No. 683184). FDE additionally acknowledges funding through the ERC Advanced grant 695671 ``QUENCH'' and support by the Science and Technology Facilities Council (STFC). {PFW acknowledges the support of the fellowship from the East Asian Core Observatories Association.}

\bibliography{ms.v17}

\begin{thebibliography}{}
\expandafter\ifx\csname natexlab\endcsname\relax\def\natexlab#1{#1}\fi
\providecommand{\url}[1]{\href{#1}{#1}}

\bibitem[{{Aumer} {et~al.}(2010){Aumer}, {Burkert}, {Johansson}, \&
  {Genzel}}]{Aumer10}
{Aumer}, M., {Burkert}, A., {Johansson}, P.~H., \& {Genzel}, R. 2010, \apj,
  719, 1230

\bibitem[{{Bacchini} {et~al.}(2019){Bacchini}, {Fraternali}, {Pezzulli},
  {Marasco}, {Iorio}, \& {Nipoti}}]{Bacchini19}
{Bacchini}, C., {Fraternali}, F., {Pezzulli}, G., {et~al.} 2019, \aap, 632,
  A127

\bibitem[{Bassett {et~al.}(2014)Bassett, Glazebrook, Fisher, Green, Wisnioski,
  Obreschkow, Cooper, Abraham, Damjanov, \& McGregor}]{Bassett14}
Bassett, R., Glazebrook, K., Fisher, D.~B., {et~al.} 2014, Monthly Notices of
  the Royal Astronomical Society, 442, 3206.
\newblock \url{https://doi.org/10.1093/mnras/stu1029}

\bibitem[{{Bekiaris} {et~al.}(2016){Bekiaris}, {Glazebrook}, {Fluke}, \&
  {Abraham}}]{Bekiaris16}
{Bekiaris}, G., {Glazebrook}, K., {Fluke}, C.~J., \& {Abraham}, R. 2016,
  \mnras, 455, 754

\bibitem[{{Bezanson} {et~al.}(2018{\natexlab{a}}){Bezanson}, {van der Wel},
  {Pacifici}, {Noeske}, {Bari{\v{s}}i{\'c}}, {Bell}, {Brammer}, {Calhau},
  {Chauke}, {van Dokkum}, {Franx}, {Gallazzi}, {van Houdt}, {Labb{\'e}},
  {Maseda}, {Mu{\~n}os-Mateos}, {Muzzin}, {van de Sand e}, {Sobral},
  {Straatman}, \& {Wu}}]{Bezanson18a}
{Bezanson}, R., {van der Wel}, A., {Pacifici}, C., {et~al.} 2018{\natexlab{a}},
  \apj, 858, 60

\bibitem[{{Bezanson} {et~al.}(2018{\natexlab{b}}){Bezanson}, {van der Wel},
  {Straatman}, {Pacifici}, {Wu}, {Bari{\v{s}}i{\'c}}, {Bell}, {Conroy},
  {D'Eugenio}, {Franx}, {Gallazzi}, {van Houdt}, {Maseda}, {Muzzin}, {van de
  Sande}, {Sobral}, \& {Spilker}}]{Bezanson18b}
{Bezanson}, R., {van der Wel}, A., {Straatman}, C., {et~al.}
  2018{\natexlab{b}}, \apjl, 868, L36

\bibitem[{{Binney} \& {Tremaine}(2008)}]{Binney08}
{Binney}, J., \& {Tremaine}, S. 2008, {Galactic Dynamics: Second Edition}

\bibitem[{{Bournaud} {et~al.}(2009){Bournaud}, {Elmegreen}, \&
  {Martig}}]{Bournaud09}
{Bournaud}, F., {Elmegreen}, B.~G., \& {Martig}, M. 2009, \apjl, 707, L1

\bibitem[{{Bournaud} {et~al.}(2010){Bournaud}, {Elmegreen}, {Teyssier},
  {Block}, \& {Puerari}}]{Bournaud10}
{Bournaud}, F., {Elmegreen}, B.~G., {Teyssier}, R., {Block}, D.~L., \&
  {Puerari}, I. 2010, \mnras, 409, 1088

\bibitem[{{Buchner} {et~al.}(2014){Buchner}, {Georgakakis}, {Nandra}, {Hsu},
  {Rangel}, {Brightman}, {Merloni}, {Salvato}, {Donley}, \&
  {Kocevski}}]{Buchner14}
{Buchner}, J., {Georgakakis}, A., {Nandra}, K., {et~al.} 2014, \aap, 564, A125

\bibitem[{{Burkert} {et~al.}(2010){Burkert}, {Genzel}, {Bouch{\'e}}, {Cresci},
  {Khochfar}, {Sommer-Larsen}, {Sternberg}, {Naab}, {F{\"o}rster Schreiber},
  {Tacconi}, {Shapiro}, {Hicks}, {Lutz}, {Davies}, {Buschkamp}, \&
  {Genel}}]{Burkert10}
{Burkert}, A., {Genzel}, R., {Bouch{\'e}}, N., {et~al.} 2010, \apj, 725, 2324

\bibitem[{{Capak} {et~al.}(2007){Capak}, {Aussel}, {Ajiki}, {McCracken},
  {Mobasher}, {Scoville}, {Shopbell}, {Taniguchi}, {Thompson}, {Tribiano},
  {Sasaki}, {Blain}, {Brusa}, {Carilli}, {Comastri}, {Carollo}, {Cassata},
  {Colbert}, {Ellis}, {Elvis}, {Giavalisco}, {Green}, {Guzzo}, {Hasinger},
  {Ilbert}, {Impey}, {Jahnke}, {Kartaltepe}, {Kneib}, {Koda}, {Koekemoer},
  {Komiyama}, {Leauthaud}, {Le Fevre}, {Lilly}, {Liu}, {Massey}, {Miyazaki},
  {Murayama}, {Nagao}, {Peacock}, {Pickles}, {Porciani}, {Renzini}, {Rhodes},
  {Rich}, {Salvato}, {Sanders}, {Scarlata}, {Schiminovich}, {Schinnerer},
  {Scodeggio}, {Sheth}, {Shioya}, {Tasca}, {Taylor}, {Yan}, \&
  {Zamorani}}]{Capak07}
{Capak}, P., {Aussel}, H., {Ajiki}, M., {et~al.} 2007, \apjs, 172, 99

\bibitem[{Cappellari(2002)}]{Cappellari02}
Cappellari, M. 2002, Monthly Notices of the Royal Astronomical Society, 333,
  400.
\newblock \url{https://doi.org/10.1046/j.1365-8711.2002.05412.x}

\bibitem[{Cappellari(2008)}]{Cappellari08}
---. 2008, Monthly Notices of the Royal Astronomical Society, 390, 71.
\newblock \url{https://doi.org/10.1111/j.1365-2966.2008.13754.x}

\bibitem[{{Cappellari}(2017)}]{Cappellari17}
{Cappellari}, M. 2017, \mnras, 466, 798

\bibitem[{{Cappellari} \& {Emsellem}(2004)}]{Cappellari04}
{Cappellari}, M., \& {Emsellem}, E. 2004, \pasp, 116, 138

\bibitem[{Cappellari {et~al.}(2006)Cappellari, Bacon, Bureau, Damen, Davies,
  De~Zeeuw, Emsellem, Falc{\'o}n-Barroso, Krajnovic, Kuntschner, McDermid,
  Peletier, Sarzi, Van Den~Bosch, \& Van De~Ven}]{Cappellari06}
Cappellari, M., Bacon, R., Bureau, M., {et~al.} 2006, Monthly Notices of the
  Royal Astronomical Society, 366, 1126.
\newblock \url{https://doi.org/10.1111/j.1365-2966.2005.09981.x}

\bibitem[{{Ceverino} {et~al.}(2010){Ceverino}, {Dekel}, \&
  {Bournaud}}]{Ceverino10}
{Ceverino}, D., {Dekel}, A., \& {Bournaud}, F. 2010, \mnras, 404, 2151

\bibitem[{{Chabrier}(2003)}]{Chabrier03}
{Chabrier}, G. 2003, \pasp, 115, 763

\bibitem[{{Cresci} {et~al.}(2009){Cresci}, {Hicks}, {Genzel}, {Schreiber},
  {Davies}, {Bouch{\'e}}, {Buschkamp}, {Genel}, {Shapiro}, {Tacconi},
  {Sommer-Larsen}, {Burkert}, {Eisenhauer}, {Gerhard}, {Lutz}, {Naab},
  {Sternberg}, {Cimatti}, {Daddi}, {Erb}, {Kurk}, {Lilly}, {Renzini},
  {Shapley}, {Steidel}, \& {Caputi}}]{Cresci09}
{Cresci}, G., {Hicks}, E.~K.~S., {Genzel}, R., {et~al.} 2009, \apj, 697, 115

\bibitem[{{Crespo G{\'o}mez} {et~al.}(2021){Crespo G{\'o}mez}, {Piqueras
  L{\'o}pez}, {Arribas}, {Pereira-Santaella}, {Colina}, \& {Rodr{\'\i}guez del
  Pino}}]{CrespoGomez21}
{Crespo G{\'o}mez}, A., {Piqueras L{\'o}pez}, J., {Arribas}, S., {et~al.} 2021,
  \aap, 650, A149

\bibitem[{{Davies} {et~al.}(2011){Davies}, {F{\"o}rster Schreiber}, {Cresci},
  {Genzel}, {Bouch{\'e}}, {Burkert}, {Buschkamp}, {Genel}, {Hicks}, {Kurk},
  {Lutz}, {Newman}, {Shapiro}, {Sternberg}, {Tacconi}, \& {Wuyts}}]{Davies11}
{Davies}, R., {F{\"o}rster Schreiber}, N.~M., {Cresci}, G., {et~al.} 2011,
  \apj, 741, 69

\bibitem[{{Dekel} {et~al.}(2009){Dekel}, {Sari}, \& {Ceverino}}]{Dekel09}
{Dekel}, A., {Sari}, R., \& {Ceverino}, D. 2009, \apj, 703, 785

\bibitem[{{Di Teodoro} \& {Fraternali}(2015)}]{DiTeodoro15}
{Di Teodoro}, E.~M., \& {Fraternali}, F. 2015, \mnras, 451, 3021

\bibitem[{{Dib} {et~al.}(2006){Dib}, {Bell}, \& {Burkert}}]{Dib06}
{Dib}, S., {Bell}, E., \& {Burkert}, A. 2006, \apj, 638, 797

\bibitem[{{Dutton} \& {Macci{\`o}}(2014)}]{Dutton14}
{Dutton}, A.~A., \& {Macci{\`o}}, A.~V. 2014, \mnras, 441, 3359

\bibitem[{{Eilers} {et~al.}(2019){Eilers}, {Hogg}, {Rix}, \& {Ness}}]{Eilers19}
{Eilers}, A.-C., {Hogg}, D.~W., {Rix}, H.-W., \& {Ness}, M.~K. 2019, \apj, 871,
  120

\bibitem[{{Elmegreen} \& {Burkert}(2010)}]{Elmegreen10}
{Elmegreen}, B.~G., \& {Burkert}, A. 2010, \apj, 712, 294

\bibitem[{{Epinat} {et~al.}(2010){Epinat}, {Amram}, {Balkowski}, \&
  {Marcelin}}]{Epinat10}
{Epinat}, B., {Amram}, P., {Balkowski}, C., \& {Marcelin}, M. 2010, \mnras,
  401, 2113

\bibitem[{{Epinat} {et~al.}(2012){Epinat}, {Tasca}, {Amram}, {Contini}, {Le
  F{\`e}vre}, {Queyrel}, {Vergani}, {Garilli}, {Kissler-Patig}, {Moultaka},
  {Paioro}, {Tresse}, {Bournaud}, {L{\'o}pez-Sanjuan}, \& {Perret}}]{Epinat12}
{Epinat}, B., {Tasca}, L., {Amram}, P., {et~al.} 2012, \aap, 539, A92

\bibitem[{{Foreman-Mackey} {et~al.}(2013){Foreman-Mackey}, {Hogg}, {Lang}, \&
  {Goodman}}]{ForemanMackey13}
{Foreman-Mackey}, D., {Hogg}, D.~W., {Lang}, D., \& {Goodman}, J. 2013, \pasp,
  125, 306

\bibitem[{{F{\"o}rster Schreiber} {et~al.}(2014){F{\"o}rster Schreiber},
  {Genzel}, {Newman}, {Kurk}, {Lutz}, {Tacconi}, {Wuyts}, {Bandara}, {Burkert},
  {Buschkamp}, {Carollo}, {Cresci}, {Daddi}, {Davies}, {Eisenhauer}, {Hicks},
  {Lang}, {Lilly}, {Mainieri}, {Mancini}, {Naab}, {Peng}, {Renzini}, {Rosario},
  {Shapiro Griffin}, {Shapley}, {Sternberg}, {Tacchella}, {Vergani},
  {Wisnioski}, {Wuyts}, \& {Zamorani}}]{NFS14}
{F{\"o}rster Schreiber}, N.~M., {Genzel}, R., {Newman}, S.~F., {et~al.} 2014,
  \apj, 787, 38

\bibitem[{{F{\"o}rster Schreiber} {et~al.}(2019){F{\"o}rster Schreiber},
  {{\"U}bler}, {Davies}, {Genzel}, {Wisnioski}, {Belli}, {Shimizu}, {Lutz},
  {Fossati}, {Herrera-Camus}, {Mendel}, {Tacconi}, {Wilman}, {Beifiori},
  {Brammer}, {Burkert}, {Carollo}, {Davies}, {Eisenhauer}, {Fabricius},
  {Lilly}, {Momcheva}, {Naab}, {Nelson}, {Price}, {Renzini}, {Saglia},
  {Sternberg}, {van Dokkum}, \& {Wuyts}}]{NFS19}
{F{\"o}rster Schreiber}, N.~M., {{\"U}bler}, H., {Davies}, R.~L., {et~al.}
  2019, \apj, 875, 21

\bibitem[{{Genzel} {et~al.}(2011){Genzel}, {Newman}, {Jones}, {F{\"o}rster
  Schreiber}, {Shapiro}, {Genel}, {Lilly}, {Renzini}, {Tacconi}, {Bouch{\'e}},
  {Burkert}, {Cresci}, {Buschkamp}, {Carollo}, {Ceverino}, {Davies}, {Dekel},
  {Eisenhauer}, {Hicks}, {Kurk}, {Lutz}, {Mancini}, {Naab}, {Peng},
  {Sternberg}, {Vergani}, \& {Zamorani}}]{Genzel11}
{Genzel}, R., {Newman}, S., {Jones}, T., {et~al.} 2011, \apj, 733, 101

\bibitem[{{Gnerucci} {et~al.}(2011){Gnerucci}, {Marconi}, {Cresci}, {Maiolino},
  {Mannucci}, {Calura}, {Cimatti}, {Cocchia}, {Grazian}, {Matteucci}, {Nagao},
  {Pozzetti}, \& {Troncoso}}]{Gnerucci11}
{Gnerucci}, A., {Marconi}, A., {Cresci}, G., {et~al.} 2011, \aap, 528, A88

\bibitem[{{Green} {et~al.}(2014){Green}, {Glazebrook}, {McGregor}, {Damjanov},
  {Wisnioski}, {Abraham}, {Colless}, {Sharp}, {Crain}, {Poole}, \&
  {McCarthy}}]{Green14}
{Green}, A.~W., {Glazebrook}, K., {McGregor}, P.~J., {et~al.} 2014, \mnras,
  437, 1070

\bibitem[{{Harrison} {et~al.}(2017){Harrison}, {Johnson}, {Swinbank}, {Stott},
  {Bower}, {Smail}, {Tiley}, {Bunker}, {Cirasuolo}, {Sobral}, {Sharples},
  {Best}, {Bureau}, {Jarvis}, \& {Magdis}}]{Harrison17}
{Harrison}, C.~M., {Johnson}, H.~L., {Swinbank}, A.~M., {et~al.} 2017, \mnras,
  467, 1965

\bibitem[{{Haynes} \& {Giovanelli}(1984)}]{Haynes84}
{Haynes}, M.~P., \& {Giovanelli}, R. 1984, \aj, 89, 758

\bibitem[{{Immeli} {et~al.}(2004){Immeli}, {Samland}, {Westera}, \&
  {Gerhard}}]{Immeli04}
{Immeli}, A., {Samland}, M., {Westera}, P., \& {Gerhard}, O. 2004, \apj, 611,
  20

\bibitem[{{Johnson} {et~al.}(2019){Johnson}, {Leja}, {Conroy}, \&
  {Speagle}}]{Johnson19}
{Johnson}, B.~D., {Leja}, J.~L., {Conroy}, C., \& {Speagle}, J.~S. 2019,
  {Prospector: Stellar population inference from spectra and SEDs}, , ,
  ascl:1905.025

\bibitem[{{Johnson} {et~al.}(2018){Johnson}, {Harrison}, {Swinbank}, {Tiley},
  {Stott}, {Bower}, {Smail}, {Bunker}, {Sobral}, {Turner}, {Best}, {Bureau},
  {Cirasuolo}, {Jarvis}, {Magdis}, {Sharples}, {Bland-Hawthorn}, {Catinella},
  {Cortese}, {Croom}, {Federrath}, {Glazebrook}, {Sweet}, {Bryant}, {Goodwin},
  {Konstantopoulos}, {Lawrence}, {Medling}, {Owers}, \& {Richards}}]{Johnson18}
{Johnson}, H.~L., {Harrison}, C.~M., {Swinbank}, A.~M., {et~al.} 2018, \mnras,
  474, 5076

\bibitem[{{Kassin} {et~al.}(2012){Kassin}, {Weiner}, {Faber}, {Gardner},
  {Willmer}, {Coil}, {Cooper}, {Devriendt}, {Dutton}, {Guhathakurta}, {Koo},
  {Metevier}, {Noeske}, \& {Primack}}]{Kassin12}
{Kassin}, S.~A., {Weiner}, B.~J., {Faber}, S.~M., {et~al.} 2012, The
  Astrophysical Journal, 758, 106

\bibitem[{{Kretschmer} {et~al.}(2021){Kretschmer}, {Dekel}, {Freundlich},
  {Lapiner}, {Ceverino}, \& {Primack}}]{Kretschmer21}
{Kretschmer}, M., {Dekel}, A., {Freundlich}, J., {et~al.} 2021, \mnras, 503,
  5238

\bibitem[{{Kriek} {et~al.}(2015){Kriek}, {Shapley}, {Reddy}, {Siana}, {Coil},
  {Mobasher}, {Freeman}, {de Groot}, {Price}, {Sanders}, {Shivaei}, {Brammer},
  {Momcheva}, {Skelton}, {van Dokkum}, {Whitaker}, {Aird}, {Azadi}, {Kassis},
  {Bullock}, {Conroy}, {Dav{\'e}}, {Kere{\v{s}}}, \& {Krumholz}}]{Kriek15}
{Kriek}, M., {Shapley}, A.~E., {Reddy}, N.~A., {et~al.} 2015, \apjs, 218, 15

\bibitem[{{Law} {et~al.}(2009){Law}, {Steidel}, {Erb}, {Larkin}, {Pettini},
  {Shapley}, \& {Wright}}]{Law09}
{Law}, D.~R., {Steidel}, C.~C., {Erb}, D.~K., {et~al.} 2009, \apj, 697, 2057

\bibitem[{{Le F{\`e}vre} {et~al.}(2003){Le F{\`e}vre}, {Saisse}, {Mancini},
  {Brau-Nogue}, {Caputi}, {Castinel}, {D'Odorico}, {Garilli}, {Kissler-Patig},
  {Lucuix}, {Mancini}, {Pauget}, {Sciarretta}, {Scodeggio}, {Tresse}, \&
  {Vettolani}}]{LeFevre03}
{Le F{\`e}vre}, O., {Saisse}, M., {Mancini}, D., {et~al.} 2003, in Society of
  Photo-Optical Instrumentation Engineers (SPIE) Conference Series, Vol. 4841,
  Instrument Design and Performance for Optical/Infrared Ground-based
  Telescopes, ed. M.~{Iye} \& A.~F.~M. {Moorwood}, 1670--1681

\bibitem[{{Le F{\`e}vre} {et~al.}(2015){Le F{\`e}vre}, {Tasca}, {Cassata},
  {Garilli}, {Le Brun}, {Maccagni}, {Pentericci}, {Thomas}, {Vanzella},
  {Zamorani}, {Zucca}, {Amorin}, {Bardelli}, {Capak}, {Cassar{\`a}},
  {Castellano}, {Cimatti}, {Cuby}, {Cucciati}, {de la Torre}, {Durkalec},
  {Fontana}, {Giavalisco}, {Grazian}, {Hathi}, {Ilbert}, {Lemaux}, {Moreau},
  {Paltani}, {Ribeiro}, {Salvato}, {Schaerer}, {Scodeggio}, {Sommariva},
  {Talia}, {Taniguchi}, {Tresse}, {Vergani}, {Wang}, {Charlot}, {Contini},
  {Fotopoulou}, {L{\'o}pez-Sanjuan}, {Mellier}, \& {Scoville}}]{LeFevre15}
{Le F{\`e}vre}, O., {Tasca}, L.~A.~M., {Cassata}, P., {et~al.} 2015, \aap, 576,
  A79

\bibitem[{{Leja} {et~al.}(2017){Leja}, {Johnson}, {Conroy}, {van Dokkum}, \&
  {Byler}}]{Leja17}
{Leja}, J., {Johnson}, B.~D., {Conroy}, C., {van Dokkum}, P.~G., \& {Byler}, N.
  2017, \apj, 837, 170

\bibitem[{{Leja} {et~al.}(2019){Leja}, {Johnson}, {Conroy}, {van Dokkum},
  {Speagle}, {Brammer}, {Momcheva}, {Skelton}, {Whitaker}, {Franx}, \&
  {Nelson}}]{Leja19}
{Leja}, J., {Johnson}, B.~D., {Conroy}, C., {et~al.} 2019, \apj, 877, 140

\bibitem[{{Leung} {et~al.}(2018){Leung}, {Leaman}, {van de Ven}, {Lyubenova},
  {Zhu}, {Bolatto}, {Falc{\'o}n-Barroso}, {Blitz}, {Dannerbauer}, {Fisher},
  {Levy}, {Sanchez}, {Utomo}, {Vogel}, {Wong}, \& {Ziegler}}]{Leung18}
{Leung}, G. Y.~C., {Leaman}, R., {van de Ven}, G., {et~al.} 2018, \mnras, 477,
  254

\bibitem[{{Levy} {et~al.}(2018){Levy}, {Bolatto}, {Teuben}, {S{\'a}nchez},
  {Barrera-Ballesteros}, {Blitz}, {Colombo}, {Garc{\'\i}a-Benito},
  {Herrera-Camus}, {Husemann}, {Kalinova}, {Lan}, {Leung}, {Mast}, {Utomo},
  {van de Ven}, {Vogel}, \& {Wong}}]{Levy18}
{Levy}, R.~C., {Bolatto}, A.~D., {Teuben}, P., {et~al.} 2018, \apj, 860, 92

\bibitem[{{Mac Low} \& {Klessen}(2004)}]{MacLow04}
{Mac Low}, M.-M., \& {Klessen}, R.~S. 2004, Reviews of Modern Physics, 76, 125

\bibitem[{{McCracken} {et~al.}(2012){McCracken}, {Milvang-Jensen}, {Dunlop},
  {Franx}, {Fynbo}, {Le F{\`e}vre}, {Holt}, {Caputi}, {Goranova}, {Buitrago},
  {Emerson}, {Freudling}, {Hudelot}, {L{\'o}pez-Sanjuan}, {Magnard}, {Mellier},
  {M{\o}ller}, {Nilsson}, {Sutherland}, {Tasca}, \& {Zabl}}]{McCracken12}
{McCracken}, H.~J., {Milvang-Jensen}, B., {Dunlop}, J., {et~al.} 2012, \aap,
  544, A156

\bibitem[{{Meidt} {et~al.}(2018){Meidt}, {Leroy}, {Rosolowsky}, {Kruijssen},
  {Schinnerer}, {Schruba}, {Pety}, {Blanc}, {Bigiel}, {Chevance}, {Hughes},
  {Querejeta}, \& {Usero}}]{Meidt18}
{Meidt}, S.~E., {Leroy}, A.~K., {Rosolowsky}, E., {et~al.} 2018, \apj, 854, 100

\bibitem[{{Meidt} {et~al.}(2020){Meidt}, {Glover}, {Kruijssen}, {Leroy},
  {Rosolowsky}, {Hughes}, {Schinnerer}, {Schruba}, {Usero}, {Bigiel}, {Blanc},
  {Chevance}, {Pety}, {Querejeta}, \& {Utomo}}]{Meidt20}
{Meidt}, S.~E., {Glover}, S. C.~O., {Kruijssen}, J.~M.~D., {et~al.} 2020, \apj,
  892, 73

\bibitem[{{Miller} {et~al.}(2011){Miller}, {Bundy}, {Sullivan}, {Ellis}, \&
  {Treu}}]{Miller11}
{Miller}, S.~H., {Bundy}, K., {Sullivan}, M., {Ellis}, R.~S., \& {Treu}, T.
  2011, \apj, 741, 115

\bibitem[{{Muzzin} {et~al.}(2013){Muzzin}, {Marchesini}, {Stefanon}, {Franx},
  {Milvang-Jensen}, {Dunlop}, {Fynbo}, {Brammer}, {Labb{\'e}}, \& {van
  Dokkum}}]{Muzzin13}
{Muzzin}, A., {Marchesini}, D., {Stefanon}, M., {et~al.} 2013, \apjs, 206, 8

\bibitem[{{Navarro} {et~al.}(1996){Navarro}, {Frenk}, \& {White}}]{Navarro96}
{Navarro}, J.~F., {Frenk}, C.~S., \& {White}, S. D.~M. 1996, \apj, 462, 563

\bibitem[{{Newman} {et~al.}(2013){Newman}, {Cooper}, {Davis}, {Faber}, {Coil},
  {Guhathakurta}, {Koo}, {Phillips}, {Conroy}, {Dutton}, {Finkbeiner}, {Gerke},
  {Rosario}, {Weiner}, {Willmer}, {Yan}, {Harker}, {Kassin}, {Konidaris},
  {Lai}, {Madgwick}, {Noeske}, {Wirth}, {Connolly}, {Kaiser}, {Kirby},
  {Lemaux}, {Lin}, {Lotz}, {Luppino}, {Marinoni}, {Matthews}, {Metevier}, \&
  {Schiavon}}]{Newman13}
{Newman}, J.~A., {Cooper}, M.~C., {Davis}, M., {et~al.} 2013, \apjs, 208, 5

\bibitem[{{Ostriker} {et~al.}(2010){Ostriker}, {McKee}, \&
  {Leroy}}]{Ostriker10}
{Ostriker}, E.~C., {McKee}, C.~F., \& {Leroy}, A.~K. 2010, \apj, 721, 975

\bibitem[{{Peng} {et~al.}(2010){Peng}, {Ho}, {Impey}, \& {Rix}}]{Peng10}
{Peng}, C.~Y., {Ho}, L.~C., {Impey}, C.~D., \& {Rix}, H.-W. 2010, \aj, 139,
  2097

\bibitem[{{Pizagno} {et~al.}(2007){Pizagno}, {Prada}, {Weinberg}, {Rix},
  {Pogge}, {Grebel}, {Harbeck}, {Blanton}, {Brinkmann}, \& {Gunn}}]{Pizagno07}
{Pizagno}, J., {Prada}, F., {Weinberg}, D.~H., {et~al.} 2007, \aj, 134, 945

\bibitem[{{Pizzella} {et~al.}(2004){Pizzella}, {Corsini}, {Vega Beltr{\'a}n},
  \& {Bertola}}]{Pizzella04}
{Pizzella}, A., {Corsini}, E.~M., {Vega Beltr{\'a}n}, J.~C., \& {Bertola}, F.
  2004, \aap, 424, 447

\bibitem[{{Price} {et~al.}(2016){Price}, {Kriek}, {Shapley}, {Reddy},
  {Freeman}, {Coil}, {de Groot}, {Shivaei}, {Siana}, {Azadi}, {Barro},
  {Mobasher}, {Sand ers}, \& {Zick}}]{Price16}
{Price}, S.~H., {Kriek}, M., {Shapley}, A.~E., {et~al.} 2016, \apj, 819, 80

\bibitem[{{Price} {et~al.}(2020){Price}, {Kriek}, {Barro}, {Shapley}, {Reddy},
  {Freeman}, {Coil}, {Shivaei}, {Azadi}, {de Groot}, {Siana}, {Mobasher}, {Sand
  ers}, {Leung}, {Fetherolf}, {Zick}, {{\"U}bler}, \& {F{\"o}rster
  Schreiber}}]{Price20}
{Price}, S.~H., {Kriek}, M., {Barro}, G., {et~al.} 2020, \apj, 894, 91

\bibitem[{{Rubin} {et~al.}(2010){Rubin}, {Weiner}, {Koo}, {Martin},
  {Prochaska}, {Coil}, \& {Newman}}]{Rubin10}
{Rubin}, K. H.~R., {Weiner}, B.~J., {Koo}, D.~C., {et~al.} 2010, \apj, 719,
  1503

\bibitem[{{Santini} {et~al.}(2017){Santini}, {Fontana}, {Castellano}, {Di
  Criscienzo}, {Merlin}, {Amorin}, {Cullen}, {Daddi}, {Dickinson}, {Dunlop},
  {Grazian}, {Lamastra}, {McLure}, {Micha{\l}owski}, {Pentericci}, \&
  {Shu}}]{Santini17}
{Santini}, P., {Fontana}, A., {Castellano}, M., {et~al.} 2017, \apj, 847, 76

\bibitem[{Scoville {et~al.}(2007)Scoville, Aussel, Brusa, Capak, Carollo,
  Elvis, Giavalisco, Guzzo, Hasinger, Impey, Kneib, LeFevre, Lilly, Mobasher,
  Renzini, Rich, Sanders, Schinnerer, Schminovich, Shopbell, Taniguchi, \&
  Tyson}]{Scoville07}
Scoville, N., Aussel, H., Brusa, M., {et~al.} 2007, The Astrophysical Journal
  Supplement Series, 172, 1.
\newblock \url{https://doi.org/10.1086/516585}

\bibitem[{{Spitzer}(1942)}]{Spitzer42}
{Spitzer}, Lyman, J. 1942, \apj, 95, 329

\bibitem[{{Stott} {et~al.}(2016){Stott}, {Swinbank}, {Johnson}, {Tiley},
  {Magdis}, {Bower}, {Bunker}, {Bureau}, {Harrison}, {Jarvis}, {Sharples},
  {Smail}, {Sobral}, {Best}, \& {Cirasuolo}}]{Stott16}
{Stott}, J.~P., {Swinbank}, A.~M., {Johnson}, H.~L., {et~al.} 2016, \mnras,
  457, 1888

\bibitem[{{Straatman} {et~al.}(2017){Straatman}, {Glazebrook}, {Kacprzak},
  {Labb{\'e}}, {Nanayakkara}, {Alcorn}, {Cowley}, {Kewley}, {Spitler}, {Tran},
  \& {Yuan}}]{Straatman17}
{Straatman}, C. M.~S., {Glazebrook}, K., {Kacprzak}, G.~G., {et~al.} 2017,
  \apj, 839, 57

\bibitem[{{Straatman} {et~al.}(2018){Straatman}, {van der Wel}, {Bezanson},
  {Pacifici}, {Gallazzi}, {Wu}, {Noeske}, {Bari{\v{s}}i{\'c}}, {Bell},
  {Brammer}, {Calhau}, {Chauke}, {Franx}, {van Houdt}, {Labb{\'e}}, {Maseda},
  {Mu{\~n}oz-Mateos}, {Muzzin}, {van de Sand e}, {Sobral}, \&
  {Spilker}}]{Straatman18}
{Straatman}, C. M.~S., {van der Wel}, A., {Bezanson}, R., {et~al.} 2018, \apjs,
  239, 27

\bibitem[{{Trujillo} {et~al.}(2001){Trujillo}, {Aguerri}, {Cepa}, \&
  {Guti{\'e}rrez}}]{Trujillo01}
{Trujillo}, I., {Aguerri}, J.~A.~L., {Cepa}, J., \& {Guti{\'e}rrez}, C.~M.
  2001, \mnras, 328, 977

\bibitem[{{Turner} {et~al.}(2017){Turner}, {Cirasuolo}, {Harrison}, {McLure},
  {Dunlop}, {Swinbank}, {Johnson}, {Sobral}, {Matthee}, \&
  {Sharples}}]{Turner17}
{Turner}, O.~J., {Cirasuolo}, M., {Harrison}, C.~M., {et~al.} 2017, \mnras,
  471, 1280

\bibitem[{{{\"U}bler} {et~al.}(2017){{\"U}bler}, {F{\"o}rster Schreiber},
  {Genzel}, {Wisnioski}, {Wuyts}, {Lang}, {Naab}, {Burkert}, {van Dokkum},
  {Tacconi}, {Wilman}, {Fossati}, {Mendel}, {Beifiori}, {Belli}, {Bender},
  {Brammer}, {Chan}, {Davies}, {Fabricius}, {Galametz}, {Lutz}, {Momcheva},
  {Nelson}, {Saglia}, {Seitz}, \& {Tadaki}}]{Ubler17}
{{\"U}bler}, H., {F{\"o}rster Schreiber}, N.~M., {Genzel}, R., {et~al.} 2017,
  \apj, 842, 121

\bibitem[{{{\"U}bler} {et~al.}(2019){{\"U}bler}, {Genzel}, {Wisnioski},
  {F{\"o}rster Schreiber}, {Shimizu}, {Price}, {Tacconi}, {Belli}, {Wilman},
  {Fossati}, {Mendel}, {Davies}, {Beifiori}, {Bender}, {Brammer}, {Burkert},
  {Chan}, {Davies}, {Fabricius}, {Galametz}, {Herrera-Camus}, {Lang}, {Lutz},
  {Momcheva}, {Naab}, {Nelson}, {Saglia}, {Tadaki}, {van Dokkum}, \&
  {Wuyts}}]{Ubler19}
{{\"U}bler}, H., {Genzel}, R., {Wisnioski}, E., {et~al.} 2019, \apj, 880, 48

\bibitem[{{van der Wel} {et~al.}(2016){van der Wel}, {Noeske}, {Bezanson},
  {Pacifici}, {Gallazzi}, {Franx}, {Mu{\~n}oz-Mateos}, {Bell}, {Brammer},
  {Charlot}, {Chauk{\'e}}, {Labb{\'e}}, {Maseda}, {Muzzin}, {Rix}, {Sobral},
  {van de Sande}, {van Dokkum}, {Wild}, \& {Wolf}}]{vdWel16}
{van der Wel}, A., {Noeske}, K., {Bezanson}, R., {et~al.} 2016, \apjs, 223, 29

\bibitem[{{van der Wel} {et~al.}(2021){van der Wel}, {Bezanson}, {D'Eugenio},
  {Straatman}, {Franx}, {van Houdt}, {Maseda}, {Gallazzi}, {Wu}, {Pacifici},
  {Barisic}, {Brammer}, {Munoz-Mateos}, {Vervalcke}, {Zibetti}, {Sobral}, {de
  Graaff}, {Calhau}, {Kaushal}, {Muzzin}, {Bell}, \& {van Dokkum}}]{vdWel21}
{van der Wel}, A., {Bezanson}, R., {D'Eugenio}, F., {et~al.} 2021, \apjs, 256,
  44

\bibitem[{{van Houdt} {et~al.}(2021){van Houdt}, {van der Wel}, {Bezanson},
  {Franx}, {D'Eugenio}, {Barisic}, {Bell}, {Gallazzi}, {de Graaff}, {Maseda},
  {Pacifici}, {van de Sande}, {Sobral}, {Straatman}, \& {Wu}}]{vHoudt21}
{van Houdt}, J., {van der Wel}, A., {Bezanson}, R., {et~al.} 2021, arXiv
  e-prints, arXiv:2108.08142

\bibitem[{{Varidel} {et~al.}(2019){Varidel}, {Croom}, {Lewis}, {Brewer}, {Di
  Teodoro}, {Bland -Hawthorn}, {Bryant}, {Federrath}, {Foster}, {Glazebrook},
  {Goodwin}, {Groves}, {Hopkins}, {Lawrence}, {L{\'o}pez-S{\'a}nchez},
  {Medling}, {Owers}, {Richards}, {Scalzo}, {Scott}, {Sweet}, {Taranu}, \& {van
  de Sande}}]{Varidel19}
{Varidel}, M.~R., {Croom}, S.~M., {Lewis}, G.~F., {et~al.} 2019, \mnras, 485,
  4024

\bibitem[{{Vergani} {et~al.}(2012){Vergani}, {Epinat}, {Contini}, {Tasca},
  {Tresse}, {Amram}, {Garilli}, {Kissler-Patig}, {Le F{\`e}vre}, {Moultaka},
  {Paioro}, {Queyrel}, \& {L{\'o}pez-Sanjuan}}]{Vergani12}
{Vergani}, D., {Epinat}, B., {Contini}, T., {et~al.} 2012, \aap, 546, A118

\bibitem[{{Weiner} {et~al.}(2006){Weiner}, {Willmer}, {Faber}, {Melbourne},
  {Kassin}, {Phillips}, {Harker}, {Metevier}, {Vogt}, \& {Koo}}]{Weiner06}
{Weiner}, B.~J., {Willmer}, C. N.~A., {Faber}, S.~M., {et~al.} 2006, \apj, 653,
  1027

\bibitem[{{Wellons} {et~al.}(2020){Wellons}, {Faucher-Gigu{\`e}re},
  {Angl{\'e}s-Alc{\'a}zar}, {Hayward}, {Feldmann}, {Hopkins}, \&
  {Kere{\v{s}}}}]{Wellons20}
{Wellons}, S., {Faucher-Gigu{\`e}re}, C.-A., {Angl{\'e}s-Alc{\'a}zar}, D.,
  {et~al.} 2020, \mnras, 497, 4051

\bibitem[{{Wisnioski} {et~al.}(2015){Wisnioski}, {F{\"o}rster Schreiber},
  {Wuyts}, {Wuyts}, {Bandara}, {Wilman}, {Genzel}, {Bender}, {Davies},
  {Fossati}, {Lang}, {Mendel}, {Beifiori}, {Brammer}, {Chan}, {Fabricius},
  {Fudamoto}, {Kulkarni}, {Kurk}, {Lutz}, {Nelson}, {Momcheva}, {Rosario},
  {Saglia}, {Seitz}, {Tacconi}, \& {van Dokkum}}]{Wisnioski15}
{Wisnioski}, E., {F{\"o}rster Schreiber}, N.~M., {Wuyts}, S., {et~al.} 2015,
  The Astrophysical Journal, 799, 209

\bibitem[{{Wisnioski} {et~al.}(2018){Wisnioski}, {Mendel}, {F{\"o}rster
  Schreiber}, {Genzel}, {Wilman}, {Wuyts}, {Belli}, {Beifiori}, {Bender},
  {Brammer}, {Chan}, {Davies}, {Davies}, {Fabricius}, {Fossati}, {Galametz},
  {Lang}, {Lutz}, {Nelson}, {Momcheva}, {Rosario}, {Saglia}, {Tacconi},
  {Tadaki}, {{\"U}bler}, \& {van Dokkum}}]{Wisnioski18}
{Wisnioski}, E., {Mendel}, J.~T., {F{\"o}rster Schreiber}, N.~M., {et~al.}
  2018, \apj, 855, 97

\bibitem[{{Wisnioski} {et~al.}(2019){Wisnioski}, {F{\"o}rster Schreiber},
  {Fossati}, {Mendel}, {Wilman}, {Genzel}, {Bender}, {Wuyts}, {Davies},
  {{\"U}bler}, {Bandara}, {Beifiori}, {Belli}, {Brammer}, {Chan}, {Davies},
  {Fabricius}, {Galametz}, {Lang}, {Lutz}, {Nelson}, {Momcheva}, {Price},
  {Rosario}, {Saglia}, {Seitz}, {Shimizu}, {Tacconi}, {Tadaki}, {van Dokkum},
  \& {Wuyts}}]{Wisnioski19}
{Wisnioski}, E., {F{\"o}rster Schreiber}, N.~M., {Fossati}, M., {et~al.} 2019,
  \apj, 886, 124

\end{thebibliography}

\appendix
\section{custom code for emission line dynamics}\label{app:hela}

\subsection{Basic functionality}

Our custom model is in its functionality most similar to MisFIT \citep{Price20}. Here we repeat for completeness the exact steps the code performs to generate a model emission line.

We first defined coordinates $x',y',z'$  and $\lambda'$, with the spatial coordinates relative to the presumed dynamic center. For the LEGA-C spectra this is pixel row 40{ in the $y'-$direction}. The wavelength $\lambda'$ can be given either in observed or rest frame. We generated a ``cube'' of observed spatial coordinates, where the $x'-$ and $z'-$ dimensions are a fraction of the $y'-$ dimension depending on the orientation of the galaxy. The cube is generated with a factor 3$\times$ higher spatial resolution than the data. The center of the cube has coordinates $(x',y',z')=(0,0,0)$.

We performed a coordinate transformation to obtain the intrinsic $x,y$ and $z$:
\begin{equation}
\begin{bmatrix}
x\\
y\\
z
\end{bmatrix}
=\begin{bmatrix}
\cos i & 0 & -\sin i\\
0 & 1 & 0\\
\sin i & 0  & \cos i
\end{bmatrix}\begin{bmatrix}
\cos \mathrm{\Delta PA} & \sin \mathrm{\Delta PA} & 0\\
-\sin \mathrm{\Delta PA} & \cos \mathrm{\Delta PA} & 0\\
0 & 0 & 1
\end{bmatrix}\begin{bmatrix}
x'\\
y'\\
z'\\
\end{bmatrix}
\end{equation}

As dynamic quantities we defined 
\begin{align}
v_{rot}(r)&=\frac{2}{\pi}V_{t} \arctan \frac{r}{R_t}\mathrm{;\ with\ }r=\sqrt{x^2+y^2}\\
\sigma_0&=constant
\end{align}

The line-of-sight projections are
\begin{align}
v_{rot,LOS}(r)&=-v_{rot}(r)\frac{y}{r}\sin i \\
\sigma_{LOS}&=\sigma_0/\sqrt{3}
\end{align}

We assumed an exponential light profile for the face-on projection with scale length $R_s=R_e/1.678$ and also an exponential profile for the edge-on projection with scale length $h_z=0.19R_s$:
\begin{equation}
I(r,z)=e^{-r/R_s}e^{-z/h_z}
\end{equation}

In order to expand the intensity cube into velocity space, we defined the coordinates
\begin{equation}
v'=\frac{\lambda'-\lambda_c}{\lambda_c}c
\end{equation}
with $\lambda_c$ the center of the emission line and $c$ the speed of light. $\lambda_c$ should be in observed or rest frame depending on $\lambda'$. We note that the resolution of $v'$ is dependent on $\lambda_c$ if the latter is a free variable in the fit. Again we oversampled by a factor $3$.

We obtained a $4-$D intensity cube $\mathbf{I(x,y,z,v')=I(r,z,v')}$, for which the intensity is diluted along the $v'-$direction, i.e.,
\begin{equation}
I(r,z,v')=\frac{I(r,z)}{\sqrt{2\pi\sigma_{LOS}^2}}\exp\frac{-(v'-v_{rot,LOS}(r))^2}{2\sigma_{LOS}^2}
\end{equation}

We integrated $I(r,z,v')$ along the $z'$ dimension, i.e., along the line of sight, to obtain $I(x',y',v')$. Then we convolved each slice at fixed coordinate $v'$ of the resulting cube with the observational PSF -- in this study a Moffat profile.

The convolved model cube, $I_{conv}(x',y',v')$ can be further reduced to a 2D model emission line. We rebinned along the $x'$ and $y'$ dimensions to retrieve the original resolution. We then projected a slit with width $\Delta x$ and infinite length along the $y'$ dimension, centered on $x'=0$, and selected those pixels for which $|x'|<\Delta x/2$. For each slice at fixed coordinate $x'$ we shifted the pixels in velocity space with an amount corresponding to the offset of the slice from $x'=0$ in pixels, to mimick instrumental broadening. Then we integrated over the slit and rebinned in velocity space to obtain $I_{conv}(y',v')$, where $v'$ can be projected back onto $\lambda'$-space, resulting in the model emission line.

\subsection{Post-processing specific to this study}

The model emission line can be directly fitted to the data, or further reduced to a velocity curve and dispersion profile. To avoid having to account for differences in the symmetry of the light profile of the data and to be able to use the joint pPXF fits to multiple emission lines, we chose the latter. We used \texttt{scipy.optimize.leastsq} to fit single Gaussian profiles to the flux in each row of the model emission line. As there is no noise, a fit to the model will give an unequivocal result. The Gaussians have a minimum instrumental dispersion of 34 km/s and were allowed to broaden further resulting in the equivalent of $\sigma_{pPXF}$. The centers of the Gaussians were used to obtain the model velocity curve. We note that because of this post-processing step the normalisation of the model intensity cube does not have to be considered.

\end{document}